\documentclass[twocolumn]{aastex63}
\usepackage{graphicx}
\usepackage[countmax]{subfloat}
\usepackage{booktabs}
\usepackage{mwe}
\usepackage{float}
\usepackage{amsmath}
\newcommand{\RNum}[1]{\uppercase\expandafter{\romannumeral #1\relax}}

\begin{document}
\title{Characterizing the optical nature of the blazar S5 1803+784 during its 2020 flare}

\correspondingauthor{Aditi Agarwal}
\email{aditi.agarwal@rri.res.in}

\author[0000-0003-4682-5166]{A. Agarwal}
\affiliation{Raman Research Institute, C. V. Raman Avenue, Sadashivanagar, Bengaluru - 560 080, INDIA}

\author[0000-0003-3820-0887]{Ashwani Pandey}
\affiliation{Indian Institute of Astrophysics, Block II,  Koramangala, Bangalore, India, 560034}

\author{Aykut \"Ozd\"onmez}
\affiliation{Ataturk University, Faculty of Science,  Department of Astronomy and Space Science, 25240, Yakutiye, Erzurum}

\author{Erg\"un Ege}
\affiliation{Istanbul University, Faculty of Science, Department of Astronomy and Space Sciences, 34116, Beyazıt, Istanbul, Turkey}

\author{Avik Kumar Das}
\affiliation{Raman Research Institute, C. V. Raman Avenue, Sadashivanagar, Bengaluru - 560 080, INDIA}

\author{Volkan Karakulak}
\affiliation{Ataturk University, Graduate School of Natural and Applied Sciences,  Department of Astronomy and Space Science, Yakutiye, 25240, Erzurum}

\begin{abstract}
We report the results from our study of the blazar S5 1803+784 carried out using the quasi-simultaneous $B$, $V$, $R$, and $I$ observations from May 2020 to July 2021 on 122 nights. Our observing campaign detected the historically bright optical flare during MJD 59063.5$-$MJD 59120.5. We also found the source in its brightest ($R_{mag}$= 13.617) and faintest ($R_{mag}$= 15.888) states till date. On 13 nights, covering both flaring and non-flaring periods, we searched for the intraday variability using the power-enhanced $F-$test and the nested ANOVA test. We found significant variability in 2 out of these 13 nights. However, no such variability was detected during the flaring period. From the correlation analysis, we observed that the emission in all optical bands were strongly correlated with a time lag of $\sim$ 0 days. To get insights into its dominant emission mechanisms, we generated the optical spectral energy distributions of the source on 79 nights and estimated the spectral indices by fitting the simple power law. Spectral index varied from 1.392 to 1.911 and showed significant variations with time and $R-$band magnitude. We have detected a mild bluer-when-brighter trend (BWB) during the whole monitoring period while a much stronger BWB trend during the flare. We also carried out a periodicity search using four different methods and found no significant periodicity during our observation duration. Based on the analysis during the flaring state of the source one can say that the emissions most likely originate from the jet rather than the accretion disk.
\end{abstract}

\keywords{Galaxies: active~-- BL Lacertae objects: general~-- BL Lacertae objects: individual (S5 1803+784)}

\section{Introduction}
\label{sect:intro}

BL Lacertae  objects (weak or no line profile in optical spectra) and flat-spectrum radio quasars are a violently variable subclass of blazars.
Blazars are characterized by high and variable distinctive polarisation along with strong and fast flux variability at all wavelengths from radio to $\gamma-$ray
\citep{1995PASP..107..803U, 1995ARA&A..33..163W}.
Flux variability in most blazars appears to be unpredictable and few of them have relatively reliable periodic variability \citep[e.g. OJ 287;][]{2008Natur.452..851V}.
Blazar variability is generally referred to in three groups:
Intraday variation (IDV) or microvariability for flux variations up to a few tenths of a magnitude spanning over a few minutes up to a day, short-term variation (STV) for variations exceeding one mag on a time scale of a few months, and long-term variation (LTV) for variations of a few mag over a longer timescale (months to many years). However, the trend of long-term light curves may indicate possibly longer recurrent time scales \citep[e.g.][]{ 2001A&A...374..435M, 2008AJ....135.1384G}. The mechanisms and periodicity underlying these flux variability are still open questions. Flux variability in blazars could be attributed to the shocks moving along the jet \citep{1996A&AS..120C.537M}, accretion disk instabilities \citep{1996ASPC..110...42W}, gravitational microlensing \citep{1987A&A...171...49S}, or the change of the Doppler factor owing to the emission plasma moving at relativistic speeds along a spiral path
\citep{2017Natur.552..374R}. Measuring flux and spectral changes of blazars using multiband observations is highly valuable as it provides information about the location, size, structure, and dynamics of the radiating region, cooling time scales of electrons in relativistic jets, and also test theoretical models \citep[e.g.][]{2003A&A...400..487C,2008AJ....135.1384G,2015MNRAS.450..541A}.

\begin{table*}
\centering
\caption{Observation log of S5 1803+784. The columns are: (1)the date of observations, (2) telescope used, (3) number of data points in each filter on a particular night, (4) total hours of observations in each filter. Columns 5, 6, 7, and 8 are same as columns 1, 2, 3, and 4, respectively. }
\label{tab:obs_log} 
\begin{tabular}{ccrrrrrccrrrrr} 
\hline\hline 
  Date of        & Telescope      & \multicolumn{4}{c}{Number of}  & Time & Date of        & Telescope   & \multicolumn{4}{c}{Number of}  & Time \\
observations     &     & \multicolumn{4}{c}{data points} & span & observations    &    & \multicolumn{4}{c}{data points} & span  \\
(yyyy mm dd)     &                & $B$ & $V$ & $R$ & $I$ & ($\sim$h) & (yyyy mm dd)   &  & $B$ & $V$ & $R$ & $I$  & ($\sim$h)  \\
\hline
2020-05-13 & T100 & 2 & 3 & 218 & 2 & 2.36 & 2020-10-28 & T60 & 2 & 3 & 1 & 2 & 0.09\\ 
2020-05-29 & T100 & 2 & 2 & 203 & 3 & 1.52 & 2020-11-02 & T60 & 1 & 2 & 1 & 2 & 0.09\\ 
2020-05-30 & T100 & 2 & 2 & 323 & 2 & 3.31 & 2020-11-07 & T60 & 1 & 1 & 1 & 1 & 0.06\\ 
2020-08-02 & T60 & 0 & 1 & 1 & 0 & 0.01 & 2020-11-11 & ATA50 & 0 & 0 & 2 & 0 & 0.07\\ 
2020-08-04 & T60 & 1 & 1 & 1 & 1 & 0.03 & 2020-11-22 & T100 & 0 & 0 & 93 & 0 & 2.74\\ 
2020-08-05 & T60 & 1 & 1 & 1 & 1 & 0.03 & 2020-11-24 & T60 & 3 & 3 & 3 & 3 & 0.19\\ 
2020-08-06 & T60 & 1 & 1 & 1 & 0 & 0.05 & 2020-11-25 & T60 & 3 & 3 & 3 & 3 & 0.44\\ 
2020-08-08 & T60 & 1 & 0 & 1 & 1 & 0.05 & 2020-11-26 & T60 & 2 & 2 & 3 & 3 & 0.44\\ 
2020-08-11 & T60 & 1 & 0 & 1 & 1 & 0.05 & 2020-11-28 & T60 & 3 & 2 & 3 & 3 & 0.44\\ 
2020-08-13 & T60 & 1 & 0 & 1 & 1 & 0.05 & 2020-11-29 & T60 & 1 & 2 & 0 & 0 & 0.15\\ 
2020-08-14 & T60 & 1 & 0 & 1 & 1 & 0.04 & 2020-12-18 & T60 & 2 & 2 & 3 & 3 & 0.34\\ 
2020-08-16 & T60 & 1 & 0 & 1 & 1 & 0.05 & 2020-12-20 & T60 & 3 & 3 & 3 & 3 & 0.38\\ 
2020-08-17 & T60 & 1 & 0 & 1 & 1 & 0.05 & 2020-12-24 & T60 & 3 & 3 & 2 & 3 & 0.37\\ 
2020-08-18 & T60 & 1 & 0 & 1 & 1 & 0.18 & 2020-12-25 & T60 & 3 & 3 & 2 & 3 & 0.37\\ 
2020-08-19 & T60 & 1 & 1 & 1 & 0 & 0.15 & 2020-12-26 & T60 & 3 & 3 & 2 & 3 & 0.37\\ 
2020-08-24 & T60 & 1 & 1 & 1 & 1 & 0.05 & 2021-01-04 & T60 & 0 & 1 & 1 & 1 & 0.08\\ 
2020-08-25 & T60 & 1 & 1 & 1 & 1 & 0.05 & 2021-01-06 & T60 & 3 & 2 & 3 & 3 & 0.44\\ 
2020-08-27 & T60 & 1 & 1 & 0 & 1 & 0.05 & 2021-01-21 & T60 & 3 & 3 & 3 & 3 & 0.38\\ 
2020-08-29 & T60 & 1 & 1 & 1 & 1 & 0.05 & 2021-01-22 & T60 & 3 & 3 & 3 & 3 & 0.38\\ 
2020-08-30 & T100 & 0 & 4 & 116 & 2 & 1.35 & 2021-01-23 & T60 & 3 & 3 & 3 & 3 & 0.38\\ 
2020-08-30 & T60 & 1 & 1 & 1 & 1 & 0.05 & 2021-02-01 & T60 & 2 & 2 & 3 & 3 & 0.32\\ 
2020-08-31 & T100 & 1 & 1 & 101 & 1 & 1.83 & 2021-02-11 & T100 & 0 & 0 & 79 & 0 & 2.31\\ 
2020-08-31 & T60 & 1 & 1 & 1 & 1 & 0.05 & 2021-05-07 & ATA50 & 0 & 0 & 22 & 0 & 1.5\\ 
2020-09-01 & T60 & 1 & 0 & 1 & 1 & 0.05 & 2021-05-14 & ATA50 & 0 & 0 & 86 & 0 & 3.85\\ 
2020-09-03 & T60 & 1 & 1 & 1 & 1 & 0.15 & 2021-05-15 & ATA50 & 0 & 0 & 25 & 0 & 2.44\\ 
2020-09-04 & T60 & 1 & 0 & 1 & 1 & 0.14 & 2021-05-16 & T100 & 1 & 2 & 8 & 2 & 0.48\\ 
2020-09-05 & T60 & 1 & 0 & 1 & 1 & 0.14 & 2021-05-17 & ATA50 & 0 & 0 & 09 & 0 & 3.11\\ 
2020-09-06 & T60 & 1 & 0 & 1 & 1 & 0.05 & 2021-05-18 & ATA50 & 0 & 0 & 32 & 0 & 2.11\\ 
2020-09-07 & T60 & 1 & 0 & 1 & 1 & 0.05 & 2021-05-19 & ATA50 & 0 & 0 & 43 & 0 & 3.88\\ 
2020-09-08 & T60 & 1 & 0 & 1 & 1 & 0.05 & 2021-06-03 & T60 & 1 & 1 & 1 & 1 & 0.11\\ 
2020-09-10 & T100 & 7 & 11 & 13 & 8 & 2.9 & 2021-06-04 & T60 & 1 & 1 & 1 & 1 & 0.08\\ 
2020-09-10 & T60 & 1 & 0 & 1 & 1 & 0.05 & 2021-06-05 & T60 & 1 & 1 & 1 & 1 & 0.08\\ 
2020-09-11 & T60 & 1 & 1 & 1 & 1 & 0.06 & 2021-06-06 & T60 & 1 & 1 & 1 & 1 & 0.08\\ 
2020-09-12 & T100 & 3 & 3 & 5 & 3 & 0.45 & 2021-06-07 & T60 & 1 & 1 & 1 & 1 & 0.08\\ 
2020-09-12 & T60 & 1 & 1 & 1 & 1 & 0.05 & 2021-06-10 & T60 & 1 & 1 & 1 & 1 & 0.07\\ 
2020-09-13 & T100 & 2 & 3 & 4 & 3 & 0.36 & 2021-06-11 & T60 & 1 & 1 & 1 & 1 & 0.07\\ 
2020-09-13 & T60 & 1 & 1 & 1 & 1 & 0.05 & 2021-06-12 & T60 & 1 & 1 & 1 & 1 & 0.07\\ 
2020-09-16 & T60 & 1 & 1 & 1 & 1 & 0.06 & 2021-06-13 & T60 & 1 & 1 & 1 & 1 & 0.07\\ 
2020-09-17 & T60 & 1 & 1 & 1 & 1 & 0.06 & 2021-06-14 & T60 & 1 & 1 & 1 & 1 & 0.08\\ 
2020-09-18 & T60 & 1 & 1 & 1 & 1 & 0.06 & 2021-06-18 & T60 & 1 & 1 & 1 & 1 & 0.08\\ 
2020-09-19 & T60 & 1 & 1 & 1 & 1 & 0.05 & 2021-06-19 & T60 & 1 & 1 & 1 & 1 & 0.08\\ 
2020-09-25 & T60 & 1 & 0 & 1 & 1 & 0.05 & 2021-06-22 & T60 & 2 & 2 & 2 & 2 & 0.22\\ 
2020-09-28 & T60 & 1 & 1 & 1 & 1 & 0.06 & 2021-06-23 & T60 & 2 & 2 & 2 & 2 & 0.22\\ 
\hline
\end{tabular}
\end{table*}

\begin{table*}
\centering
\caption{Table 1 cntd.}
\label{tab:obs_log1} 
\begin{tabular}{ccrrrrrccrrrrr} 
\hline\hline 
  Date of        & Telescope      & \multicolumn{4}{c}{Number of}  & Time & Date of        & Telescope   & \multicolumn{4}{c}{Number of}  & Time \\
observations     &     & \multicolumn{4}{c}{data points} & span & observations    &    & \multicolumn{4}{c}{data points} & span  \\
(yyyy mm dd)     &                & $B$ & $V$ & $R$ & $I$ & ($\sim$h) & (yyyy mm dd)   &  & $B$ & $V$ & $R$ & $I$  & ($\sim$h)  \\
\hline
2020-09-30 & T60 & 1 & 1 & 1 & 1 & 0.06 & 2021-06-25 & T60 & 2 & 2 & 2 & 2 & 0.22\\ 
2020-10-02 & T60 & 0 & 1 & 1 & 1 & 0.03 & 2021-06-28 & T60 & 2 & 2 & 2 & 2 & 0.22\\ 
2020-10-03 & T60 & 1 & 1 & 1 & 1 & 0.06 & 2021-07-01 & T60 & 2 & 2 & 2 & 2 & 0.22\\ 
2020-10-04 & T60 & 1 & 1 & 1 & 1 & 0.06 & 2021-07-05 & T60 & 2 & 2 & 2 & 2 & 0.27\\ 
2020-10-06 & T60 & 0 & 1 & 1 & 1 & 0.03 & 2021-07-06 & T60 & 2 & 2 & 2 & 2 & 0.27\\ 
2020-10-07 & T60 & 1 & 0 & 0 & 0 & 0 & 2021-07-07 & T60 & 2 & 2 & 2 & 2 & 0.22\\ 
2020-10-09 & T60 & 1 & 1 & 1 & 1 & 0.05 & 2021-07-08 & T60 & 2 & 2 & 2 & 2 & 0.22\\ 
2020-10-10 & T60 & 2 & 2 & 2 & 2 & 0.29 & 2021-07-09 & T60 & 1 & 1 & 1 & 1 & 0.08\\ 
2020-10-11 & T60 & 2 & 2 & 2 & 2 & 0.29 & 2021-07-10 & T60 & 1 & 1 & 1 & 1 & 0.08\\ 
2020-10-12 & T100 & 3 & 4 & 4 & 3 & 0.54 & 2021-07-11 & T60 & 1 & 1 & 0 & 1 & 0.04\\ 
2020-10-12 & T60 & 2 & 2 & 2 & 2 & 0.29 & 2021-07-12 & T60 & 1 & 1 & 1 & 1 & 0.08\\ 
2020-10-13 & T60 & 2 & 2 & 2 & 2 & 0.29 & 2021-07-13 & T60 & 2 & 1 & 2 & 2 & 0.23\\ 
2020-10-15 & T60 & 2 & 2 & 2 & 2 & 0.29 & 2021-07-14 & T60 & 1 & 2 & 2 & 2 & 0.17\\ 
2020-10-17 & T60 & 2 & 2 & 2 & 2 & 0.29 & 2021-07-15 & T60 & 2 & 2 & 2 & 2 & 0.23\\ 
2020-10-22 & T60 & 0 & 1 & 1 & 1 & 0.03 & 2021-07-17 & T60 & 1 & 1 & 1 & 1 & 0.08\\ 
2020-10-24 & T60 & 2 & 2 & 2 & 2 & 0.12 & 2021-07-20 & T60 & 1 & 1 & 1 & 1 & 0.07\\ 
2020-10-25 & T60 & 0 & 0 & 0 & 2 & 0.06 & 2021-07-21 & T60 & 1 & 1 & 1 & 1 & 0.08\\ 
2020-10-27 & T60 & 2 & 2 & 2 & 1 & 0.12 & 2021-07-30 & T60 & 1 & 1 & 1 & 1 & 0.07\\
\hline
\end{tabular}
\end{table*}

Emission from blazars spans the complete EM spectrum, thus allowing us to study them on a wider frequency range extending up to very high energy $\gamma-$rays.
Blazars display a double-humped structure in their broadband spectral energy distributions \citep[SEDs, e.g.][]{1997MNRAS.289..136F}.
The first hump, which is the low energy component, peaks in the Infra-red (IR) to X-rays, while the second (high energy) hump peaks from GeV to TeV frequencies. The synchrotron emission from the relativistic particles associated with the relativistic jets attributes to the low energy part
of the SED, while according to the widely accepted leptonic scenarios, the high energy hump could be due to the inverse-Compton (IC) scattering of low-frequency photons by the highly energetic particles. However, the origin of the latter component is still a question. Various models have been proposed to understand the high energy features \citep[e.g.][]{2007Ap&SS.307...69B}. The non-thermal leptonic processes which can explain the second component include the synchrotron self-Compton (SSC) model and the external Compton (EC) model.
According to the SSC model \citep[e.g.][and references therein]{2002PASA...19..138M}, the IC emission from the population of electrons up-scattering the low energy photons is responsible for the high energy features of the SED. In contrast, according to the EC model, the second hump is due to the photons from the ambient medium, e.g., accretion disk, broad-line region, and dusty torus \citep{1993ApJ...416..458D, 1994ApJS...90..923S}. In hadronic scenarios, very high energy $\gamma$-rays are produced by a variety of mechanisms such as photo-pion and photo-pair interactions, which also generate neutrinos in the process \citep{1993A&A...269...67M}. Both leptonic and hadronic models have shown great success in representing SEDs except in certain scenarios where leptonic models may pose some problems for the observed data.

\begin{figure}
\centering
\includegraphics[width=\columnwidth,clip=true]{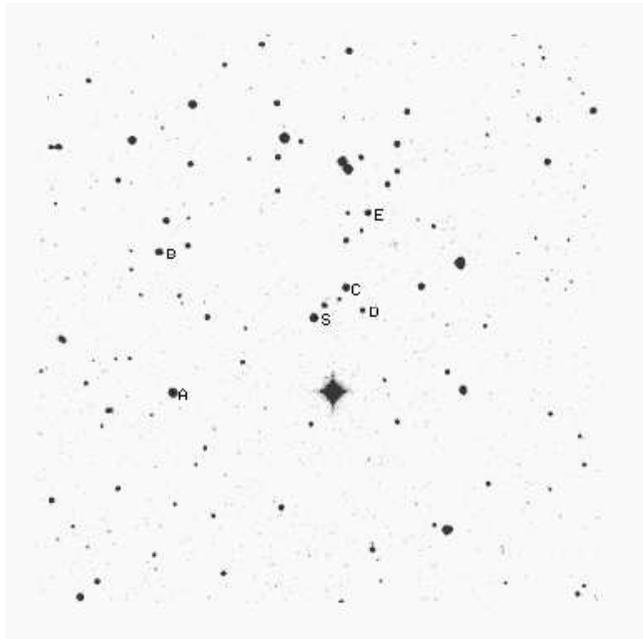}
\caption{Finding chart of S5 1803+784 where S denotes the blazar and A, B, C, D, E are the standard stars observed from the frame.}
\label{fig:chart}
\end{figure}

Blazars are further classified based on the location of the synchrotron peak ($\nu_{\rm syn}$) into three sub-classes: low synchrotron frequency peaked (LSP) with $\nu_{\rm syn} \leq$ 10$^{14}$ Hz, intermediate synchrotron frequency peaked (ISP) have 10$^{14} < \nu_{\rm syn} <$ 10$^{15}$ Hz, and high synchrotron frequency peaked (HSP) $\nu_{\rm syn} \geq$ 10$^{15}$ Hz \citep{2010ApJ...716...30A}.
 \citet{2016ApJS..226...20F} studied a large sample of blazars and slightly revised the above classification scheme of blazars. According to the authors, LSPs have the same location of their synchrotron peak but ISPs lie in the range 10$^{14} < \nu_{\rm syn} <$ 10$^{15.3}$ Hz, whereas the HSPs will have $\nu_{\rm syn} \geq$ 10$^{15.3}$ Hz. In addition to these subclasses of blazars, another growing sub-class have come to light which is popularly known as extreme high frequency peaked BL Lacertae objects (EHBLs), with the synchrotron peak frequency ranging from medium to hard, i.e., $\nu_{\rm syn}$ lying at  $> $ 1 keV  \citep[or $ > 10^{17}$ Hz;][]{2001A&A...371..512C} and therefore EHBLs are considered to be good candidates for TeV studies. 

S5 1803+784 is a BL Lacertae object with a redshift z=0.680 \citep{1996ApJS..107..541L} and a large optical polarisation \citep{2013ApJ...772...14C}. S5 1803+784 is a source that has been observed in radio bands and high energy bands while occasionally in optical bands.
It is characterised by strong X-ray and $\gamma$-ray radiation and strong flux variation the entire electromagnetic spectrum
\citep[see][]{2018MNRAS.478..359K, 2002AJ....124...53N, 2012AcPol..52a..39N, 2021MNRAS.502.6177N}.
The only systematic optical monitoring was made by \citet{2002AJ....124...53N, 2012AcPol..52a..39N, 2021MNRAS.502.6177N} between 1996-2021. This long-term monitoring revealed strong flaring activities with large variability. At the same time, analyzing optical color index vs. flux variations displayed minor variations.
A recent study by \citet{2021MNRAS.502.6177N} demonstrates the lack of any strong correlation among multi-band emissions (i.e., radio-optical, optical-X ray, and X-ray-$\gamma$ ray fluxes). A periodic oscillation of the relativistic jet with a period of 6 years was claimed by \citet{2018MNRAS.478..359K} based on analysis of VLBI images. However, \citet{2002AJ....124...53N, 2012AcPol..52a..39N, 2021MNRAS.502.6177N} did not find any periodicity in their long-term optical observations.

Since a one-zone SSC model appears too simple to explain the source behavior, \citet{2018MNRAS.478..359K} suggested an IC origin for the high-energy emission of the source, nearly co-spatial with the optical region, and radio components originating from the core moves outwards, with ejection epochs compatible with those of the two largest $\gamma$-ray flares. AGNs have diverse radiation mechanisms dominant in various regions; thus, long-term multi-wavelength observations are extremely helpful in understanding the complete and detailed picture.

Our fundamental motivation is to analyze the flux and spectral variability of the source in optical regime on intraday to longer timescales. In this paper, we present the results of quasi-simultaneous optical observations from 2020 May to 2021 July. We captured the source at its historic maxima in BVRI passbands. We also investigated the periodicity in the source on longer timescales along with inter-band correlation. Moreover, we generated the SEDs during 2020$-$2021.

The paper is structured as follows: In Section 2, we outline the properties of the telescopes used for data acquisition along with the data reduction procedure followed. In Section 3, we describe several analysis methods such as power enhanced F-test, nested ANOVA, and variability amplitude. We present our results in Section 4 and finally, discuss the different models and emission mechanisms responsible for the source behavior in Section 5.

\section{Observations and data reduction}
\label{sect:obs}

To investigate the optical properties of the blazar S5 1803+784, we performed quasi-simultaneous observations of the source in $BVRI$ from May 2020 to July 2021 for 122 nights and collected a total of $\sim$ 2100 $BVRI$ frames.

During our observations, three telescopes were used, namely, 1.0\,m RC (T100) telescope, 60\,cm RC robotic (T60) telescope, and 0.5m RC (ATA50) telescope in Turkey.
The technical details of 1.0\,m RC telescope and 60\,cm RC robotic telescope are summarized in
\citet{2021A&A...645A.137A} while 0.5m f/8 RC (ATA50) telescope is located at Ataturk University, Erzurum, and it is equipped with Apogee Alta U230 2K CCD with a field of view of $26.3\arcmin\times26.3\arcmin$, and standard Johnson BVR and SDSS ugriz filters. The log of our photometric observations, along with the total period of observations on a particular night, is listed in Tables\,\ref{tab:obs_log} and \ref{tab:obs_log1}.
Photometric images of the target field were taken in the B, V, R, and I passbands in an alternative sequence. Depending upon the sky condition and brightness of the source, the exposure time ranges from 30$-$300 seconds.

\begin{figure*}
\centering
\includegraphics[width=\linewidth,clip=true]{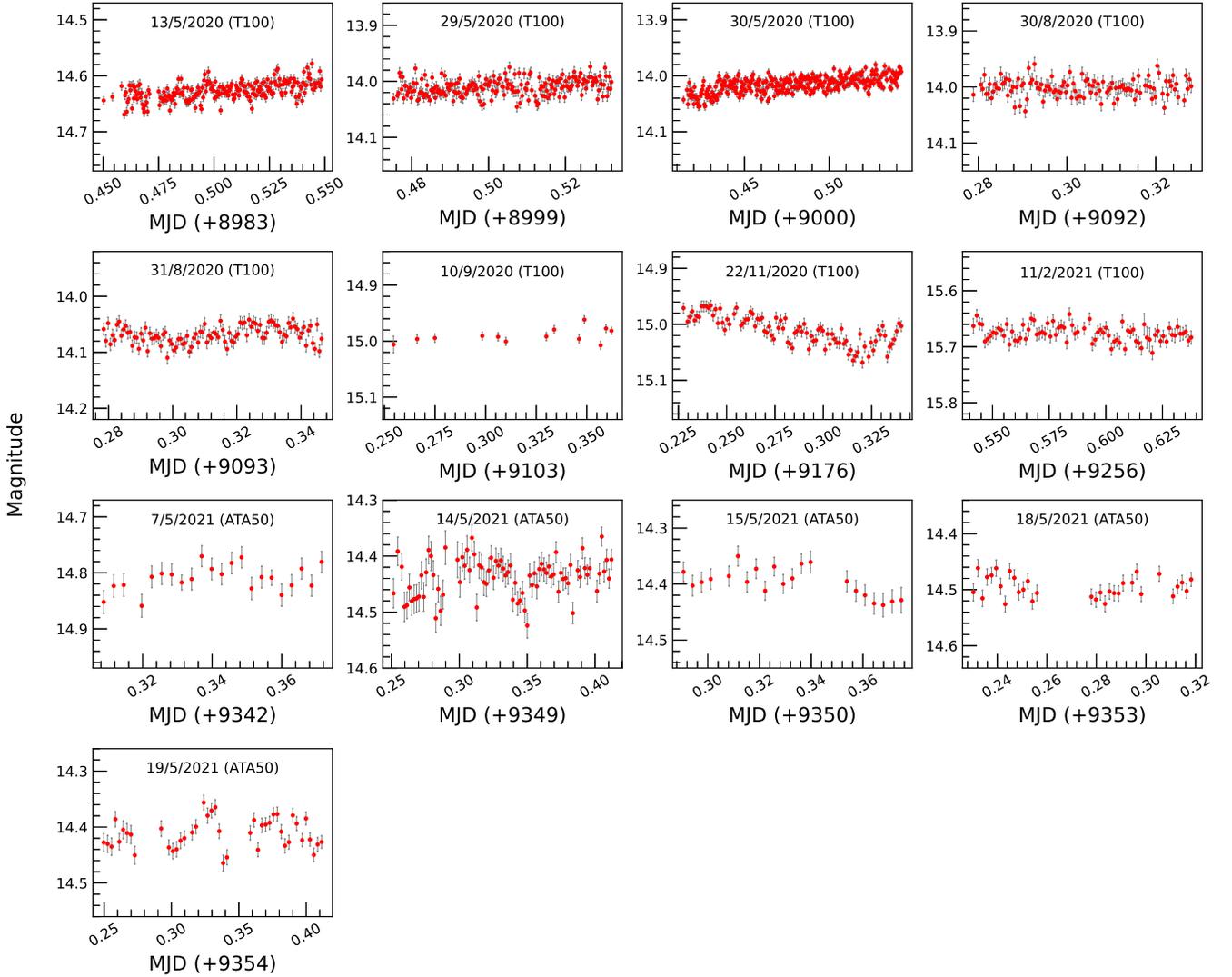}
\caption{\label{fig:lc}Optical $R-$band intraday light curves for S5 1803+784. The date of observation and the telescope code are written in the top of each plot.}
\end{figure*}

We used a similar standard data reduction methodology as in \citet{2019MNRAS.488.4093A}; i.e., the bias subtraction, twilight flat-fielding, and cosmic-ray removal.
After pre-processing the raw images, aperture photometry was carried out to extract the instrumental magnitudes of the blazar and all the comparison stars in the source frame using our script developed with python and its packages.
To find the optimum aperture, we first used different concentric aperture radii as 1.0, 1.2, 1.4, 1.6, 1.8, 2.0, 2.5, and 3.0 times the full width at half-maximum (FWHM) of the stars in the field. Whereas, to subtract the background, the radii of the sky annulus was set to approximately five times the FWHM. Instrumental magnitudes of the blazar and comparison stars in the field were extracted. The field used for our analysis is displayed in
the Fig.\,\ref{fig:chart} \citep{2002AJ....124...53N}.

We used the instrumental magnitudes for 1.4 $\times$ FWHM of aperture values with the best signal to noise ratio (S/N) and minimum standard deviation of the difference between instrumental magnitudes of standard stars (differential magnitudes) for our final analysis.

As emphasized by \citet{2007MNRAS.374..357C}, significant differences between the standard stars and blazar might lead to statistically significant yet false detection of microvariability.
Keeping this in mind, for calibration of the instrumental magnitudes of the blazar, we used stars A and B from the Fig.\,\ref{fig:chart} having brightness closest to the blazar.

\section{Analysis Techniques}

To precisely quantify the variability characteristics in the optical BVRI light curves of the source on diverse timescales, we employed two of the most recent and more powerful statistical methods, namely: the power-enhanced F-test and the nested analysis of variance (ANOVA) test \citep{2014AJ....148...93D, 2015AJ....150...44D}. As pointed out by \citet{2015AJ....150...44D}, various other statistical tests such as C-, F-, and Chi-square tests, which have been widely used to search for optical variability in the past, have many limitations and caveats, thus causing less reliable results. Analyzing various simulated variable quasars, the authors pointed out that we can attain more reliable results by increasing the number of comparison stars in our analysis. Therefore, we chose the power-enhanced F-test and the ANOVA test, involving multiple comparison stars in the analysis. These two methods are briefly discussed below.

\subsection{{\bf Power-enhanced F-test}} 

The conventional F-test compares sample variance of the blazars and the non-variable standard star from the observed frame. While, in the power-enhanced F-test, we take the brightest comparison star as a reference to derive the differential light curves (DLCs) of
the blazar and rest of the comparison stars \citep[e.g.][]{2016MNRAS.460.3950P,2020MNRAS.496.1430P,2020ApJ...890...72P}.
The power-enhanced F-statistics is defined as:

\begin{equation}
F_{enh} = \frac{s_{bl}^2}{s_c^2},
\end{equation}

where $s_{bl}^2$ is the estimated differential variance of the source while $s_{src}^2$ represents the combined variance of the comparison stars calculated using Equation (2) of \citet{2019ApJ...871..192P}.

We observed two or more comparison stars with magnitudes closer to the blazar and in its proximity for this study. We selected the brightest one as the reference of these comparison stars and the remaining (k) ones as the comparison stars. Since the blazar and the comparison stars are from the same observation frame, they share the same number of observations (N) on any particular night; thus, degrees of freedom (DOF) for the above expression is given as $\nu_{bl} = N-1$ and $\nu_c = k(N-1)$, respectively. The critical F value ($F_{critical}$) at 99\% confidence level is calculated. The values derived from the F-statistic with $\nu_{bl} = N_{bl}-1$ DOF in the numerator and $\nu_c = k(N_{bl}-1)$ DOF in the denominator are compared to the $F_{critical}$ value. A particular light curve is marked as variable if $F_{enh}$ $\geq$ $F_{critical}$, otherwise non-variable.

\begin{table*}
\caption{Results of IDV analysis of S5 1803$+$78.}            
\label{tab:var_res1}                   
\centering 
\resizebox{\textwidth} {!}{                     
\begin{tabular}{lccccccccc}           
\hline\hline                		 
Obs. date & MJD &  \multicolumn{3}{c}{{\it Power-enhanced  F-test}} & \multicolumn{3}{c}{{\it Nested ANOVA}} &  Status  & Amplitude\\
\cmidrule[0.03cm](r){3-5}\cmidrule[0.03cm](r){6-8}mm-dd-yyyy &  & DoF($\nu_1$,$\nu_2$ ) & $F_{\rm enh}$ & $F_c$  & DoF($\nu_1$,$\nu_2$ ) & $F$ & $F_c$ & & $\%$ \\
\hline
5-13-2020  &  58982 & 215,215 & 1.00 & 1.37 & 	42,172 & 2.92 & 1.70  & NV  & - \\ 
5-29-2020  &  58998 & 196,196 & 1.26 & 1.40 & 	38,156 & 1.54 & 1.74  & NV  &  - \\ 
5-30-2020  &  58999 & 317,317 & 2.66 & 1.30 & 	62,252 & 4.80 & 1.55  & V   &  7.47 \\ 
8-30-2020  &  59091 & 114,114 & 1.23 & 1.55 & 	22, 92 & 0.75 & 2.04  & NV  &  - \\ 
8-31-2020  &  59092 &  99, 99 & 1.52 & 1.60 & 	19, 80 & 5.54 & 2.14  & NV  &  - \\ 
9-10-2020  &  59102 &  12, 12 & 0.71 & 4.16 & 	 1,  8 & 3.21 & 11.26 & NV  &  - \\ 
11-22-2020 &  59175 &  90, 90 & 3.70 & 1.64 &   17, 72 & 12.77 & 2.23 & V   &  10.10 \\ 
2-11-2021  &  59256 &  77, 77 & 0.77 & 1.71 & 	14, 60 & 1.46 & 2.39  & NV  &  - \\ 
5-7-2021   &  59341 &  21, 21 & 0.79 & 2.86 & 	 3, 16 & 3.44 & 5.29  & NV  &  - \\ 
5-14-2021  &  59348 &  82, 82 & 1.61 & 1.68 & 	15, 64 & 3.65 & 2.33  & NV  &  - \\ 
5-15-2021  &  59349 &  20, 20 & 1.88 & 2.94 & 	 3, 16 & 4.49 & 5.29  & NV  & - \\ 
5-18-2021  &  59352 &  31, 31 & 1.22 & 2.35 & 	 5, 24 & 0.99 & 3.90  & NV  &  - \\ 
5-19-2021  &  59353 &  42, 42 & 0.71 & 2.08 & 	 7, 32 & 2.55 & 3.26  & NV  &  - \\ 

\hline                          
\end{tabular}}
\end{table*}

\subsection{{\bf Nested ANOVA test}}

The nested ANOVA is a highly robust method to check the variability in blazars' light curves. All comparison stars from the field are used as a reference to extract the differential light curves of the target blazar \citep{2015AJ....150...44D}. ANOVA test compares the dispersion within different data groups and between the groups drawn out from the observations of a single object. Unlike the power-enhanced F test, no comparison star is used here. Thereby increasing the number of stars to be used for our analysis. The differential LCs of the blazar are grouped into different temporal groups such that each group contains five points. We then calculated the mean square due to groups ($MS_G$) and also mean square due to nested observations in groups ($MS_{O(G)}$), which are used to estimate the F-statistic 
as follows \citep[e.g.][]{2015AJ....150...44D,2019ApJ...871..192P}:
\begin{equation}
F_{} = \frac{MS_{G}}{MS_{O(G)}},
\end{equation}

Null hypothesis is rejected and the light curve is considered as variable (V) if the $F-$value exceeds the critical value $F_{\nu_{1}, \nu_{2}}^{(\alpha)}$ at a significance level of 99\% ($\alpha= 0.01$), else it is marked as non-variable (NV). Here, the two DOFs i.e., $\nu_{1}$ and $\nu_{2}$ are defined as $a - 1$ and $a(b - 1)$, respectively, where a is the number of groups and b is the total number of
observations in the sample.

Using multiple statistical tests to search for variability in AGNs can further increase the reliability of the analysis.
Therefore, to ascertain the variability characteristics of our source on intraday timescales, we have adopted the above two criteria. A light curve is marked as variable if both the tests could detect significant variability for $\alpha= 0.01$.

\subsection{\bf Amplitude of Variability}

To quantify the variation of the light curves on any given night, we used variability amplitude which is defined as follows \citet{1996A&A...305...42H}:

\begin{eqnarray}
A = \sqrt {(A_{max} - A_{min})^{2} - 2\sigma^{2}} 
\end{eqnarray}

where A$_{max}$ and A$_{min}$ are the maximum and minimum differential instrumental magnitudes in the respective blazar LC while $\sigma$ represents the mean error. 

\begin{figure*}
\centering
\includegraphics[width=1.01\linewidth,clip=true]{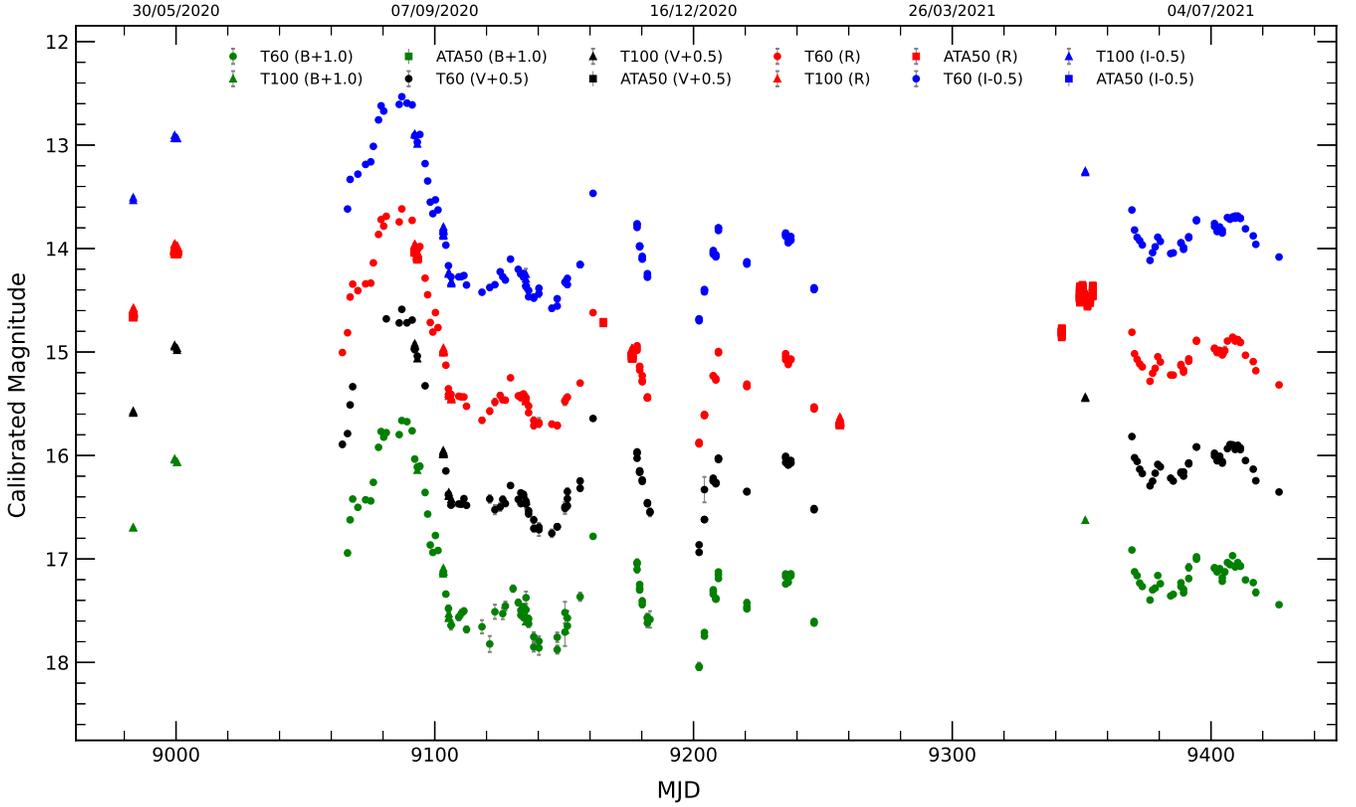}
\caption{\label{fig:stvlc}Long-term light curves of S5 1803+784 in $B$,$V$, $R$, and $I$ bands.}
\end{figure*}

\begin{table*}
\caption{Results of LTV analysis of S5 1803$+$78.}            
\label{tab:var_res2}                   
\centering 
\resizebox{\textwidth} {!}{                     
\begin{tabular}{cccccc}           
\hline\hline                		 
Band &  Brightest magnitude/MJD  &  Faintest magnitude/MJD  & Average magnitude  &   Variability amplitude (\%)\\
\hline
B  &  14.662$\pm$0.018/59086.76861  &  17.047$\pm$0.028/59201.66517  &  16.138$\pm$0.002  & 238.481 \\ 
V  &  14.088$\pm$0.011/59086.76972  &  16.436$\pm$0.013/59201.66785  &  15.568$\pm$0.002  & 234.842 \\ 
R  &  13.617$\pm$0.009/59086.77027  &  15.888$\pm$0.015/59201.65714  &  14.437$\pm$0.000  & 227.060 \\ 
I  &  13.033$\pm$0.009/59086.77073  &  15.198$\pm$0.012/59201.67119  &  14.373$\pm$0.001  & 216.566 \\ 
\hline                          
\end{tabular}}
\end{table*}

\section{Results}
\label{sect:res}

\subsection{Flux Variability}

An exhaustive and detailed search for blazar variability at different wavelengths is necessary to understand the size and/or location of the emission region and the involved particle acceleration mechanisms and radiative processes.
For IDV studies, we considered only those nights when observations were carried out for a minimum of 1 hr so that we have enough photometric data points to detect variability. We have 13 IDV nights with more than 1 hr of observations in the $R-$band following this criteria. Visual inspection of these intraday light curves hints towards the presence of variability on some occasions. Therefore, to statistically characterize IDV during these 13 nights, we used the power-enhanced F-test and the ANOVA test as described in Section 3. Of these 13 nights, the blazar S5 1803$+$78 was variable on only two nights, i.e., May 30, 2020, and November 22, 2020, with amplitude of variability reaching 7.47 and 10.10, respectively. During the rest of the 11 nights, light curves show no or very few  fluctuations. The results of IDV analysis are summarized in Table\,\ref{tab:var_res1}. The calibrated optical R band IDV LCs of our source S5 1803$+$78  are shown in Figure \,\ref{fig:lc}.

We observed the blazar from May 2020 to July 2021, large enough to
search for long-term variations (LTV). The optical $B-$, $V-$, $R-$, and $I-$ band LTV LCs of our source during the above mentioned period are shown in the Fig.\,\ref{fig:stvlc}.
We have shifted the B-, V- and I-band LCs w.r.t. R band by +1.0, +0.5, and -0.5 mag, respectively, to make the long-term light curves more clearly visible. Offsets applied for the same are mentioned in the plot.

On the months to yearly timescales, the optical $BVRI$ light curves of the blazar showed large flux variations.
Based on the optical data spanning the period from May 2020 to July 2021, we found a flaring period for the blazar S5 1803$+$78.
As depicted in Fig.\,\ref{fig:stvlc}, the flare started from $\sim$ MJD 59063.5 (August 2, 2020) and ended around MJD 59120.5 (September 28, 2020) spanning a period of 57 days.
The blazar S5 1803$+$78 was in the brightest state on August 25, 2020, B$_{mag}$ = 14.662, V$_{mag}$ = 14.088, R$_{mag}$ = 13.617, and I$_{mag}$ = 13.033 while the faintest state of the source was detected on December 18, 2020, with $B$, $V$, $R$, and $I$ brightness values of 17.047, 16.436, 15.888, and 15.198, respectively. The faintest flux level as reported by \citet{2021MNRAS.502.6177N} was on March 18, 2015, with the flux in R band $\sim$ 15.0 magnitude whereas we reported the R band flux of 15.888 which even fainter by $\sim$ 0.89 mag.
The maximum flux state identified for this source has been reported by \citet{2021MNRAS.502.6177N} with the R band magnitude of 13.7 on December 06, 2016. Therefore, the flare observed during our monitoring period on August 25, 2020, marks the brightest state of the blazar S5 1803$+$78 with an R band magnitude of 13.617.
For our entire monitoring period, we also estimated the average $BVRI$ magnitudes of S5 1803+784, which are as follows: the average B magnitude was 16.138, V magnitude was 15.568, R band magnitude was 14.437, and the I band magnitude was 14.373. The percentage of variation in the amplitude of the source over the entire monitoring period was found to be 238.481\%, 234.842\%, 227.060\%, and 216.566\% for B, V, R, I, respectively.
Above results are summarized in the Table\,\ref{tab:var_res2}.
Most
interestingly, data analysed here do not show strong minute scale variations even during the flaring state of the target.
On the other hand, large amplitude variations are found on STV/LTV timescales thereby implying that the observed variability trend could be governed by emission regions of size similar to STV/LTV timescales.

\subsection{Cross-correlation Analysis and Periodicity search}

To quantify any correlations between the optical-optical emission and to estimate the possible lags, we used the discrete correlation function (DCF) technique which was first introduced by \citet{1988ApJ...333..646E} and later used by many authors \citep[][and references therein]{2007A&A...469..899H, 2015MNRAS.450..541A, 2017ApJ...841..123P, 2021MNRAS.504.1427A}. One of the advantages of DCF is that it accounts for the irregular sampling of the data set. When the light curves are compared with themselves, we get discrete auto-correlations (DACFs). DACF of each light curve is used to determine if the peaks are obtained due to the lag between different frequencies or indicate the presence of quasi-periodicity within each light curve.

\begin{figure}
\centering
\includegraphics[width=8.5cm, height=6.5cm]{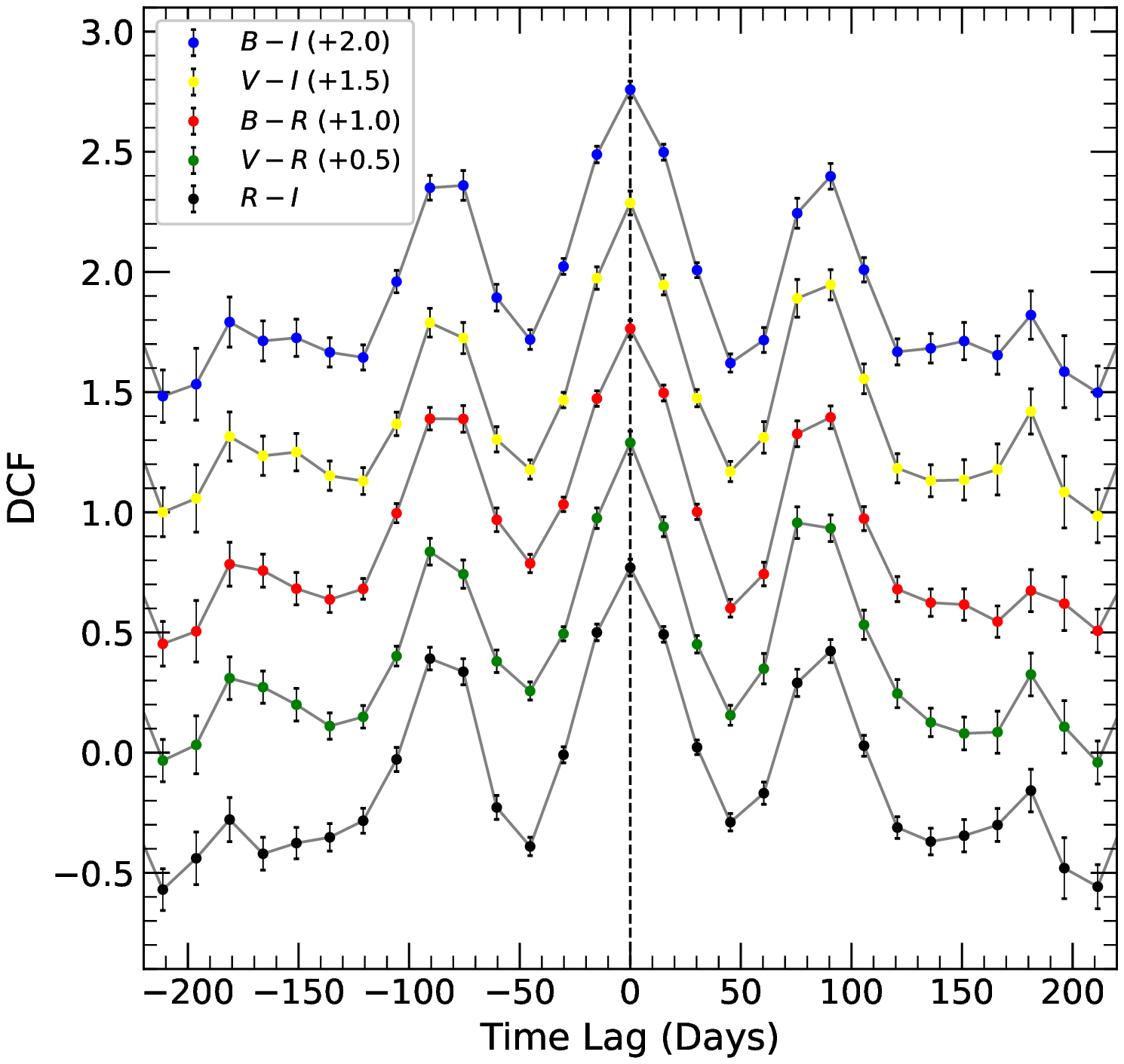}
\includegraphics[width=8.5cm, height=6.5cm]{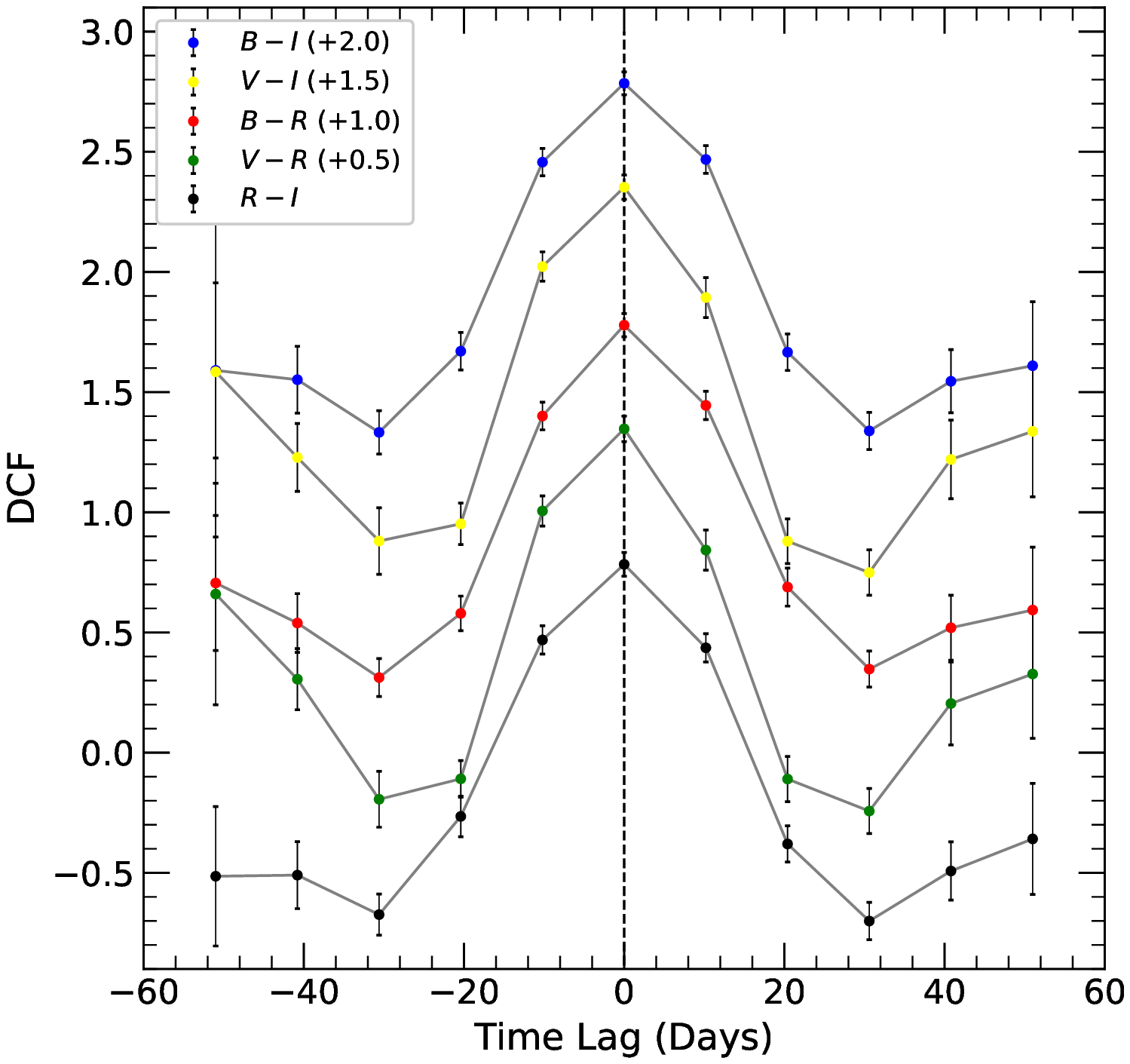}
\caption{\label{fig:dcf_fig}Cross-correlation analysis for $B$,$V$, $R$, and $I$ bands using discrete correlation function for the entire monitoring period (left) and for the flaring period (right).}

\end{figure}

To consider the effects due to densely sampled intraday light curves, we nightly binned the data set. In this regard, we obtained the magnitudes for those nights by taking the weighted average. We also took the mean of corresponding MJDs.
Finally, we performed DCF analysis on various optical-optical light curves for the entire observational period. The plots for the same are displayed in Fig.\,\ref{fig:dcf_fig}. We found a strong correlation between all possible combinations of the optical light curves with a time delay at $\sim$ 0 days. The above result was dominant for the whole observation period and also during the enhanced activity in the source during the period, i.e., $\sim$ MJD 59063.5 (August 2, 2020) to MJD 59120.5 (September 28, 2020). Optical DCFs during the flare are shown in the Fig.\,\ref{fig:dcf_fig}.

\begin{figure}
\centering
\includegraphics[width=8.5cm, height=6.5cm]{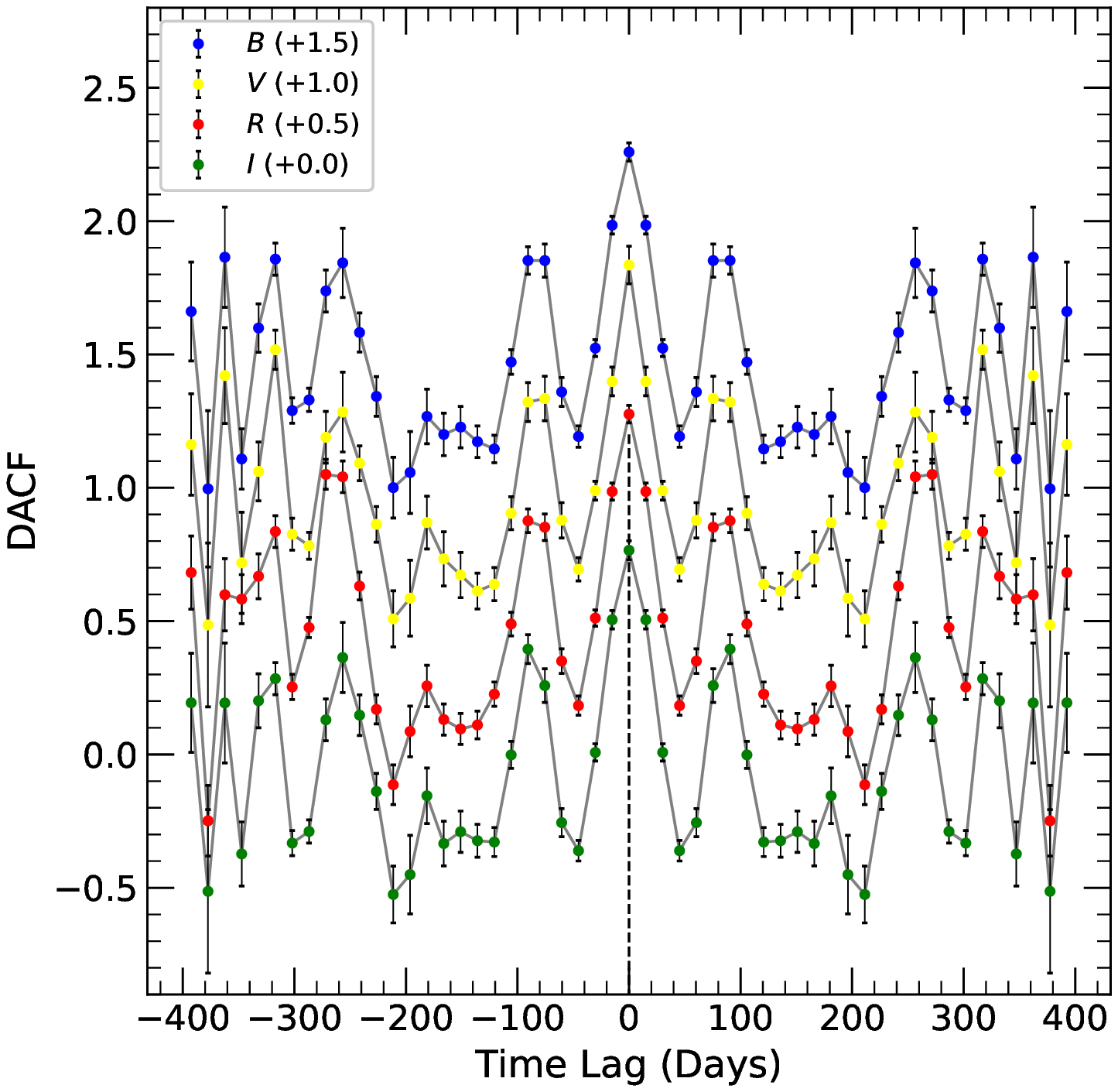}
\includegraphics[width=8.5cm, height=6.5cm]{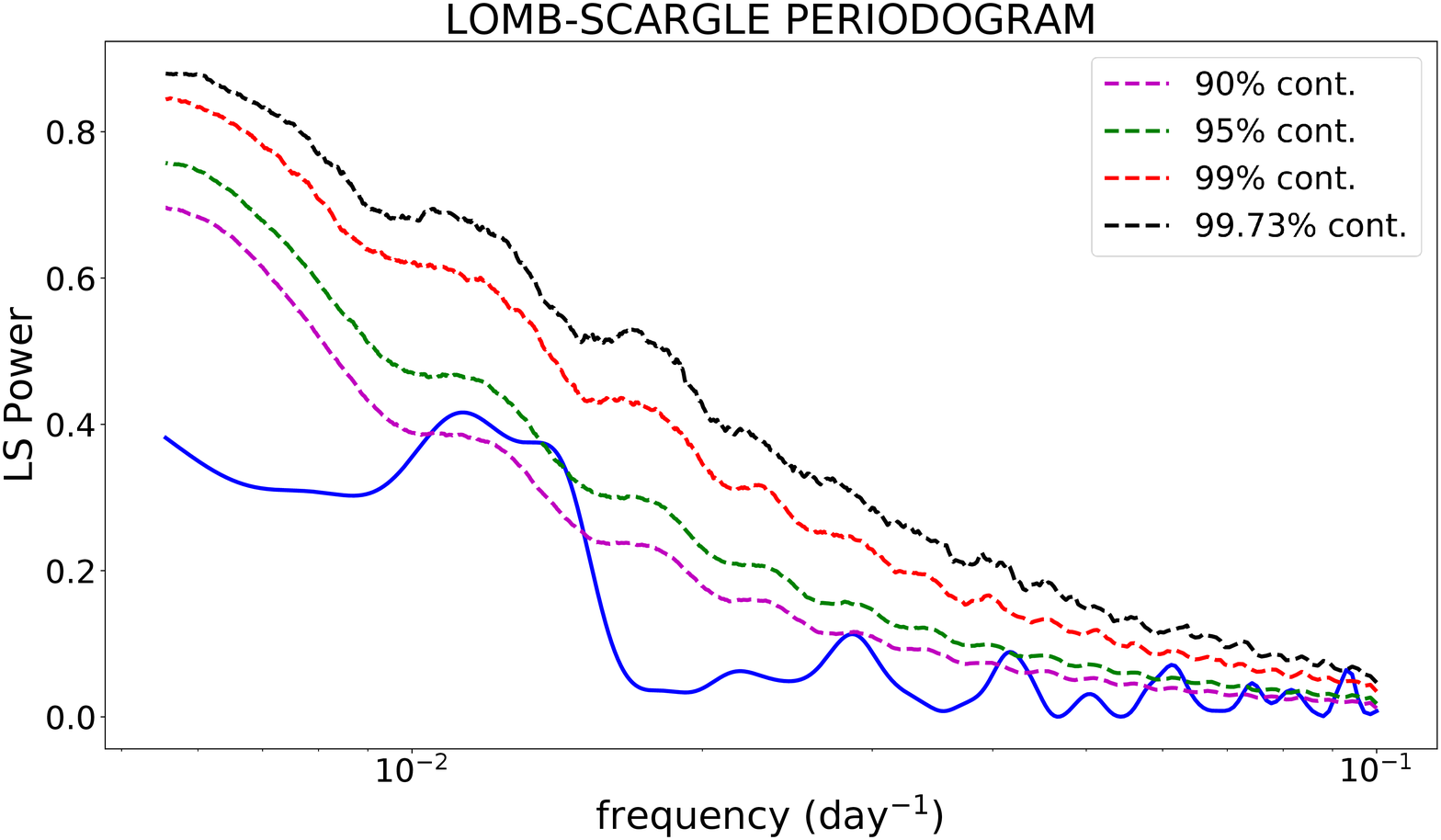}
\includegraphics[width=8.5cm, height=6.5cm]{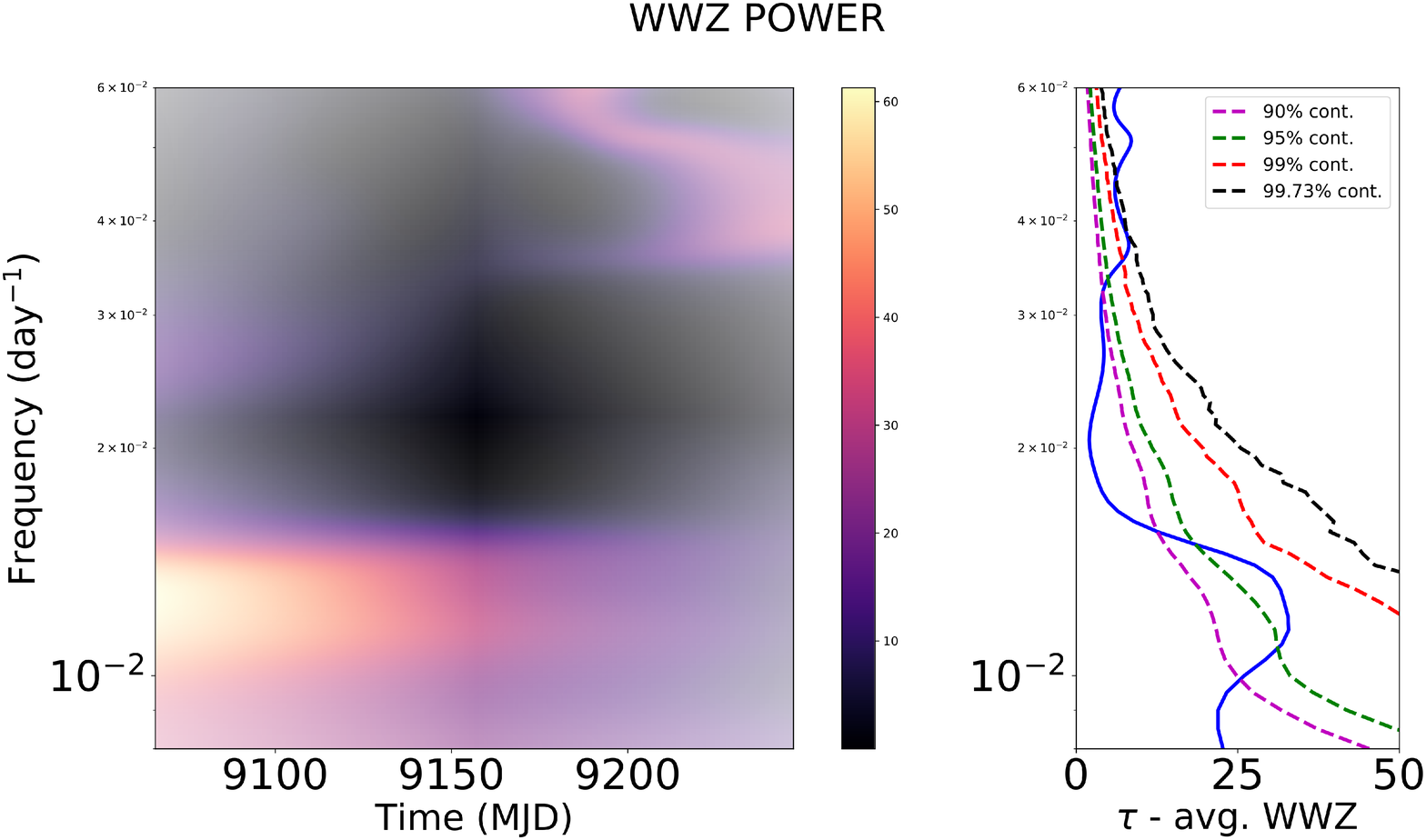}
\caption{\label{fig:period_fig}DACF for $B$,$V$, $R$, and $I$ bands (top left), LSP for B band (top right), and WWZ for the B band (bottom) for the entire monitoring period.}
\end{figure}

To search for quasi-periodicity in the source, we used a large number of analysis techniques \citep[][and references therein]{1996AJ....112.1709F, 2021Galax...9...20A} e.g., structure-function (SF), DACF, Weighted Wavelet Z-transform (WWZ), and Lomb-Scargle periodogram (LSP). For DACF, we considered the detection to be significant if the DACF values were greater than 0.5.
The DACFs indicate the presence of non-zero side peaks on a timescale of a few days but with DACF values smaller than 0.5. Whereas for robust estimation of periodicity using SF, WWZ, and LSP, we generated a large number of light curves following \cite{1995A&A...300..707T}. Using these simulated light curves, we then estimated the significance of the QPO detected using the above three methods. All above tests confirm the absence of any clear periodic variability in the optical light curves of the blazar S5 1803+784. The results from the above analysis are displayed in Fig.\,\ref{fig:period_fig}.

\begin{table*}
\caption{Straight line fits to optical SEDs of blazar S5 1803+784.} 
\label{tab:sed} 
\centering 
\begin{tabular}{cccccccccc} 
\hline\hline 
MJD & $\alpha$ & $C$ & $r$ & $p$  & 	MJD & $\alpha$ & $C$ & $r$ & $p$ \\
\hline
58982.96772	&  1.609$\pm$ 0.007 & -1.715$\pm$ 0.107 & -0.995 & 4.770e-03 &	59203.71728	&  1.746$\pm$ 0.019 & -0.096$\pm$ 0.279 & -0.993 & 7.325e-03 \\
58998.98042	&  1.494$\pm$ 0.009 & -3.150$\pm$ 0.137 & -0.994 & 5.688e-03 &	59207.13390	&  1.689$\pm$ 0.010 & -0.789$\pm$ 0.147 & -1.000 & 7.659e-05 \\
58999.92728	&  1.520$\pm$ 0.008 & -2.774$\pm$ 0.120 & -0.995 & 4.932e-03 &	59208.13391	&  1.707$\pm$ 0.007 & -0.539$\pm$ 0.097 & -0.999 & 6.084e-04 \\
59065.79383	&  1.687$\pm$ 0.033 & -0.647$\pm$ 0.480 & -0.998 & 2.163e-03 &	59209.13391	&  1.773$\pm$ 0.009 &  0.543$\pm$ 0.128 & -0.999 & 5.381e-04 \\
59066.79105	&  1.651$\pm$ 0.020 & -1.047$\pm$ 0.294 & -0.998 & 2.042e-03 &	59220.13778	&  1.734$\pm$ 0.021 & -0.166$\pm$ 0.303 & -1.000 & 4.028e-04 \\
59085.77261	&  1.513$\pm$ 0.016 & -2.777$\pm$ 0.239 & -0.999 & 1.151e-03 &	59235.11852	&  1.682$\pm$ 0.015 & -0.807$\pm$ 0.215 & -0.998 & 2.148e-03 \\
59086.76983	&  1.394$\pm$ 0.025 & -4.469$\pm$ 0.373 & -0.997 & 3.412e-03 &	59236.11852	&  1.638$\pm$ 0.012 & -1.474$\pm$ 0.169 & -0.999 & 1.202e-03 \\
59090.76682	&  1.427$\pm$ 0.019 & -4.026$\pm$ 0.283 & -0.998 & 1.681e-03 &	59237.11852	&  1.629$\pm$ 0.007 & -1.600$\pm$ 0.108 & -0.999 & 5.571e-04 \\
59091.76404	&  1.392$\pm$ 0.026 & -4.654$\pm$ 0.383 & -0.999 & 1.340e-03 &	59246.10932	&  1.558$\pm$ 0.020 & -2.835$\pm$ 0.296 & -0.999 & 1.141e-03 \\
59092.78255	&  1.501$\pm$ 0.029 & -3.074$\pm$ 0.420 & -0.996 & 3.809e-03 &	59350.92200	&  1.795$\pm$ 0.013 &  1.094$\pm$ 0.192 & -0.996 & 3.519e-03 \\
59095.77343	&  1.509$\pm$ 0.025 & -3.064$\pm$ 0.373 & -0.998 & 1.730e-03 &	59368.94119	&  1.679$\pm$ 0.010 & -0.765$\pm$ 0.146 & -1.000 & 3.195e-04 \\
59102.82033	&  1.672$\pm$ 0.007 & -0.934$\pm$ 0.100 & -0.997 & 2.870e-03 &	59369.95820	&  1.703$\pm$ 0.013 & -0.496$\pm$ 0.194 & -1.000 & 2.758e-04 \\
59103.79566	&  1.764$\pm$ 0.037 &  0.351$\pm$ 0.547 & -0.996 & 4.003e-03 &	59370.95542	&  1.646$\pm$ 0.021 & -1.353$\pm$ 0.307 & -0.999 & 7.234e-04 \\
59104.91954	&  1.739$\pm$ 0.015 & -0.115$\pm$ 0.223 & -0.996 & 4.331e-03 &	59371.95265	&  1.712$\pm$ 0.015 & -0.403$\pm$ 0.215 & -1.000 & 2.261e-04 \\
59105.92044	&  1.602$\pm$ 0.020 & -2.157$\pm$ 0.291 & -0.994 & 6.406e-03 &	59372.94987	&  1.697$\pm$ 0.013 & -0.634$\pm$ 0.191 & -1.000 & 3.085e-04 \\
59108.78177	&  1.693$\pm$ 0.047 & -0.817$\pm$ 0.693 & -0.999 & 9.965e-04 &	59375.94512	&  1.649$\pm$ 0.017 & -1.401$\pm$ 0.255 & -0.999 & 7.203e-04 \\
59109.77900	&  1.659$\pm$ 0.054 & -1.314$\pm$ 0.785 & -1.000 & 4.051e-04 &	59376.94234	&  1.669$\pm$ 0.017 & -1.070$\pm$ 0.243 & -1.000 & 4.414e-04 \\
59110.77622	&  1.610$\pm$ 0.043 & -2.028$\pm$ 0.624 & -1.000 & 4.779e-04 &	59377.93956	&  1.677$\pm$ 0.013 & -0.936$\pm$ 0.187 & -0.999 & 6.989e-04 \\
59111.77346	&  1.666$\pm$ 0.046 & -1.236$\pm$ 0.670 & -0.994 & 5.861e-03 &	59378.93678	&  1.647$\pm$ 0.014 & -1.331$\pm$ 0.212 & -0.999 & 6.661e-04 \\
59120.74868	&  1.589$\pm$ 0.084 & -2.371$\pm$ 1.229 & -0.971 & 2.913e-02 &	59379.93366	&  1.666$\pm$ 0.019 & -1.075$\pm$ 0.272 & -0.999 & 1.473e-03 \\
59122.74335	&  1.540$\pm$ 0.084 & -3.073$\pm$ 1.230 & -0.998 & 1.567e-03 &	59383.92279	&  1.658$\pm$ 0.016 & -1.232$\pm$ 0.237 & -0.998 & 1.686e-03 \\
59125.78247	&  1.618$\pm$ 0.064 & -1.909$\pm$ 0.940 & -0.999 & 1.095e-03 &	59384.92001	&  1.696$\pm$ 0.019 & -0.674$\pm$ 0.280 & -1.000 & 2.872e-04 \\
59126.77969	&  1.535$\pm$ 0.059 & -3.129$\pm$ 0.858 & -0.999 & 6.743e-04 &	59387.90868	&  1.708$\pm$ 0.015 & -0.463$\pm$ 0.222 & -1.000 & 2.447e-04 \\
59131.76587	&  1.633$\pm$ 0.052 & -1.674$\pm$ 0.766 & -0.998 & 2.129e-03 &	59388.90590	&  1.676$\pm$ 0.021 & -0.952$\pm$ 0.309 & -0.999 & 1.188e-03 \\
59132.77812	&  1.653$\pm$ 0.029 & -1.396$\pm$ 0.419 & -0.999 & 7.459e-04 &	59390.90034	&  1.645$\pm$ 0.019 & -1.362$\pm$ 0.280 & -1.000 & 1.207e-05 \\
59133.77535	&  1.601$\pm$ 0.032 & -2.148$\pm$ 0.475 & -0.997 & 2.548e-03 &	59393.89194	&  1.666$\pm$ 0.010 & -0.995$\pm$ 0.149 & -1.000 & 3.162e-04 \\
59134.72219	&  1.710$\pm$ 0.022 & -0.562$\pm$ 0.319 & -0.998 & 2.262e-03 &	59400.87296	&  1.732$\pm$ 0.012 & -0.043$\pm$ 0.181 & -1.000 & 1.404e-04 \\
59135.76977	&  1.509$\pm$ 0.037 & -3.557$\pm$ 0.545 & -0.999 & 8.772e-04 &	59401.87018	&  1.708$\pm$ 0.012 & -0.423$\pm$ 0.172 & -1.000 & 2.325e-04 \\
59137.76423	&  1.726$\pm$ 0.051 & -0.411$\pm$ 0.745 & -0.999 & 1.307e-03 &	59402.87035	&  1.718$\pm$ 0.010 & -0.257$\pm$ 0.141 & -1.000 & 1.381e-04 \\
59139.75867	&  1.911$\pm$ 0.065 &  2.297$\pm$ 0.953 & -1.000 & 4.625e-04 &	59403.86758	&  1.775$\pm$ 0.011 &  0.560$\pm$ 0.160 & -0.999 & 7.563e-04 \\
59146.82911	&  1.660$\pm$ 0.040 & -1.403$\pm$ 0.591 & -0.999 & 1.321e-03 &	59405.86260	&  1.758$\pm$ 0.015 &  0.365$\pm$ 0.225 & -1.000 & 2.136e-04 \\
59149.82111	&  1.647$\pm$ 0.098 & -1.502$\pm$ 1.433 & -0.999 & 1.444e-03 &	59407.85704	&  1.676$\pm$ 0.013 & -0.827$\pm$ 0.193 & -1.000 & 4.930e-04 \\
59150.75872	&  1.527$\pm$ 0.055 & -3.250$\pm$ 0.800 & -0.992 & 7.907e-03 &	59408.85369	&  1.756$\pm$ 0.015 &  0.334$\pm$ 0.220 & -0.999 & 8.112e-04 \\
59155.69414	&  1.532$\pm$ 0.049 & -3.113$\pm$ 0.714 & -0.999 & 9.812e-04 &	59409.85270	&  1.749$\pm$ 0.009 &  0.227$\pm$ 0.136 & -1.000 & 4.583e-04 \\
59160.68115	&  1.722$\pm$ 0.034 & -0.067$\pm$ 0.501 & -0.998 & 2.233e-03 &	59410.84807	&  1.757$\pm$ 0.009 &  0.343$\pm$ 0.130 & -0.999 & 6.219e-04 \\
59177.69894	&  1.704$\pm$ 0.025 & -0.463$\pm$ 0.368 & -1.000 & 6.381e-05 &	59412.84338	&  1.823$\pm$ 0.014 &  1.266$\pm$ 0.210 & -0.999 & 7.346e-04 \\
59178.70550	&  1.662$\pm$ 0.015 & -1.153$\pm$ 0.213 & -0.999 & 9.858e-04 &	59415.83563	&  1.800$\pm$ 0.017 &  0.896$\pm$ 0.253 & -1.000 & 1.723e-05 \\
59179.70116	&  1.640$\pm$ 0.022 & -1.523$\pm$ 0.323 & -0.996 & 3.865e-03 &	59416.83250	&  1.843$\pm$ 0.031 &  1.490$\pm$ 0.450 & -1.000 & 2.485e-04 \\
59181.69795	&  1.693$\pm$ 0.022 & -0.819$\pm$ 0.318 & -0.999 & 9.028e-04 &	59425.80553	&  1.834$\pm$ 0.017 &  1.315$\pm$ 0.255 & -1.000 & 7.693e-05 \\
59201.66377	&  1.735$\pm$ 0.025 & -0.366$\pm$ 0.360 & -0.999 & 1.043e-03 &			&		    &			&	 &	      \\
	
\hline 
\end{tabular}
\\
$\alpha$ = spectral index and $C$ = intercept of log($F_{\nu}$) against log($\nu$); $r$ = Correlation coefficient; $p$ = null hypothesis probability
\end{table*}

\begin{figure*}
\centering
\includegraphics[width=0.65\linewidth,clip=true]{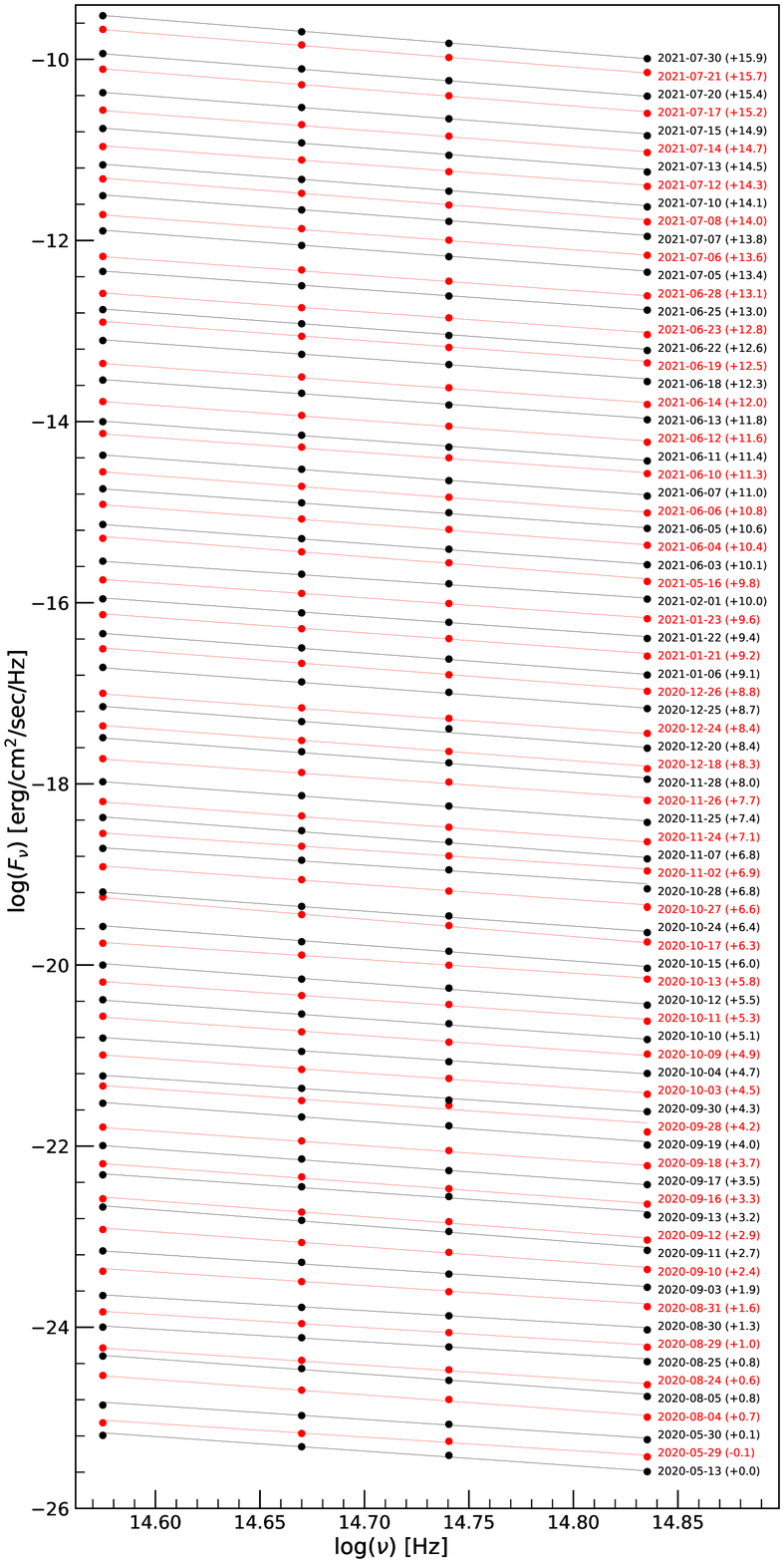}
\caption{\label{fig:sed_fig}Spectral energy distributions of S5 1803+784 in $B$,$V$, $R$, and $I$ bands.}
\end{figure*}

\subsection{Spectral energy distribution}
\label{sect:sed}
To better understand the spectral variations in our source, we extracted the multiband optical SEDs for all those dates when we have quasi-simultaneous observations in all the four bands. For this, it is necessary to correct the calibrated $BVRI$ fluxes for the Galactic extinction ($A_{\lambda}$), taken from the NASA/IPAC Extragalactic Database (NED\footnote{\url{https://ned.ipac.caltech.edu/}}): $A_{B}$ = 0.190\,mag, $A_{V}$ = 0.143\,mag, $A_{R}$ = 0.113\,mag, and $A_{I}$ =  0.079\,mag. These corrected magnitudes in B, V, R, and I bands are then converted to corresponding fluxes following \citet{1998A&A...333..231B}.
We didn’t apply any subtraction of the host galaxy component from our optical fluxes owing the following reasons: (1) the host galaxy of S5 1803+784 could not be resolved/decomposed from deep imaging data \citep{2002A&A...381..810P}. Since the blazar S5 1803+784 is a point source in our optical images, the two-dimensional decomposition such as software package Galfit \citep{2002AJ....124..266P} cannot be applied for subtraction from images; (2) the ratio between the core and the extended flux densities of S5 1805 calculated from optical and near-IR spectra \citep{2001AJ....122..565R} is only $\sim$ \%4; (3) the contribution of the host galaxy varies by the seeing (equal to aperture size), and the seeing (thus aperture) differs from 1$\arcsec$ to 3$\arcsec$ throughout our observation period. This causes an uncertainty in subtracting the host galaxy flux, typically about 10\% -- 20\% error \citep{1999PASP..111.1223N}; (4) in previous optical studies \citep{2021MNRAS.502.6177N, 2002AJ....124...53N} host galaxy subtraction was not performed as the Ca II break is undetectable or is likely negligible in the spectra published in previous studies \citep{1996ApJS..107..541L, 2001AJ....122..565R}. 
All of these reasons imply that the contribution of the host galaxy is negligible and cannot be accurately obtained from our optical data.

We have 79 nights with simultaneous $BVRI$ data sets and the optical SEDs for S5 1803+784 on those 79 nights are displayed in the Fig.\,\ref{fig:sed_fig}. The optical SEDs are well defined by simple power law ($F_{\nu} = A\nu^{\alpha}$, where $\alpha$ is the optical spectral index). To derive the spectral indices during these 79 nights, we fitted each SED with a straight line of the form: ($log(F_{\nu}) = -\alpha~log(\nu)+ C$). The results from this linear fits are given in Table\,\ref{tab:sed}. As evident from this table, the spectral index during our observation campaign varies from 1.392 to 1.911 while the weighted mean spectral index is 1.673 $\pm$ 0.002 which is in good agreement with \citet{1996ApJS..107..541L}, \citet{2021MNRAS.502.6177N}, and references therein.

\begin{figure}
\centering
\includegraphics[width=8.5cm, height=6cm]{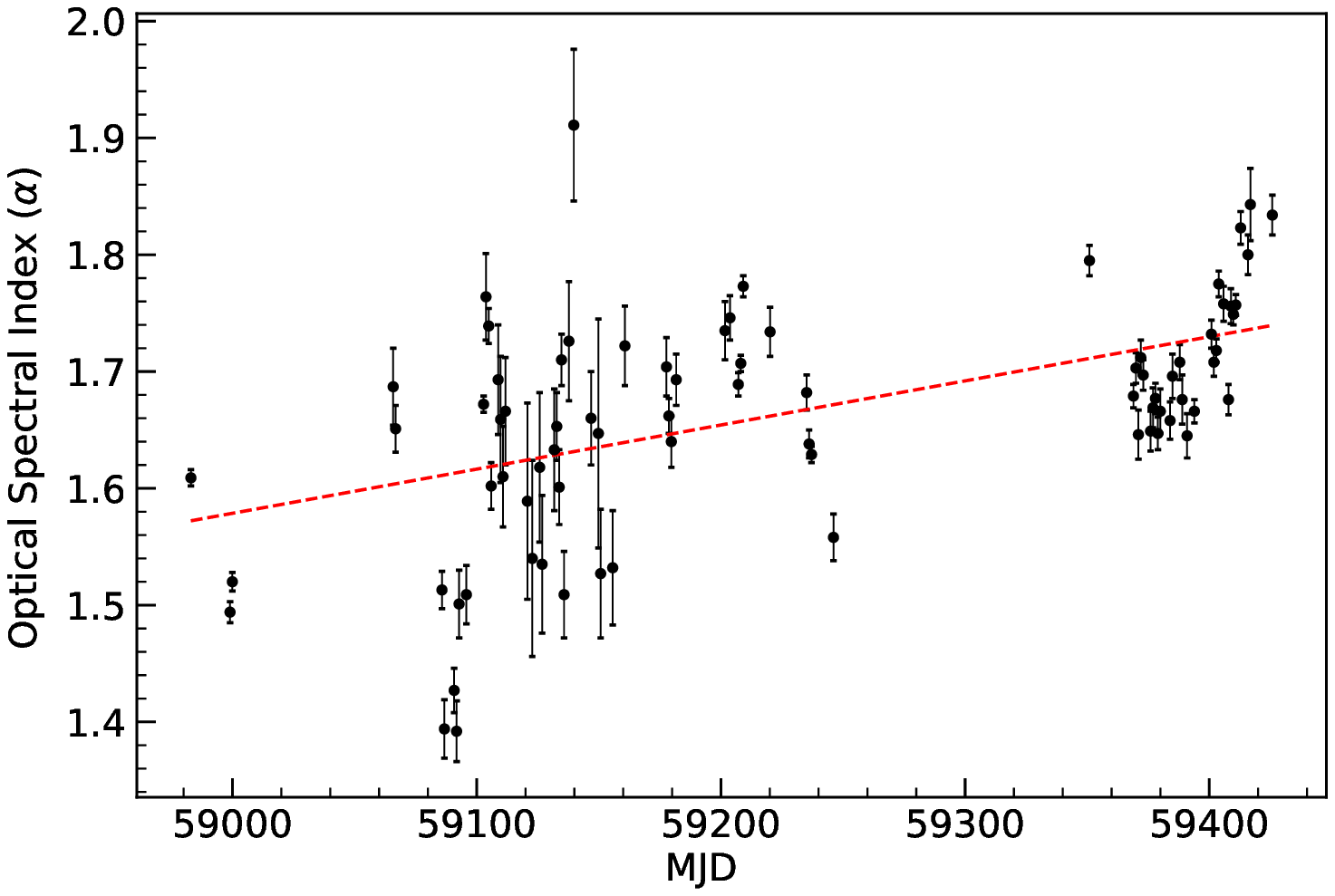}
\includegraphics[width=8.5cm, height=6cm]{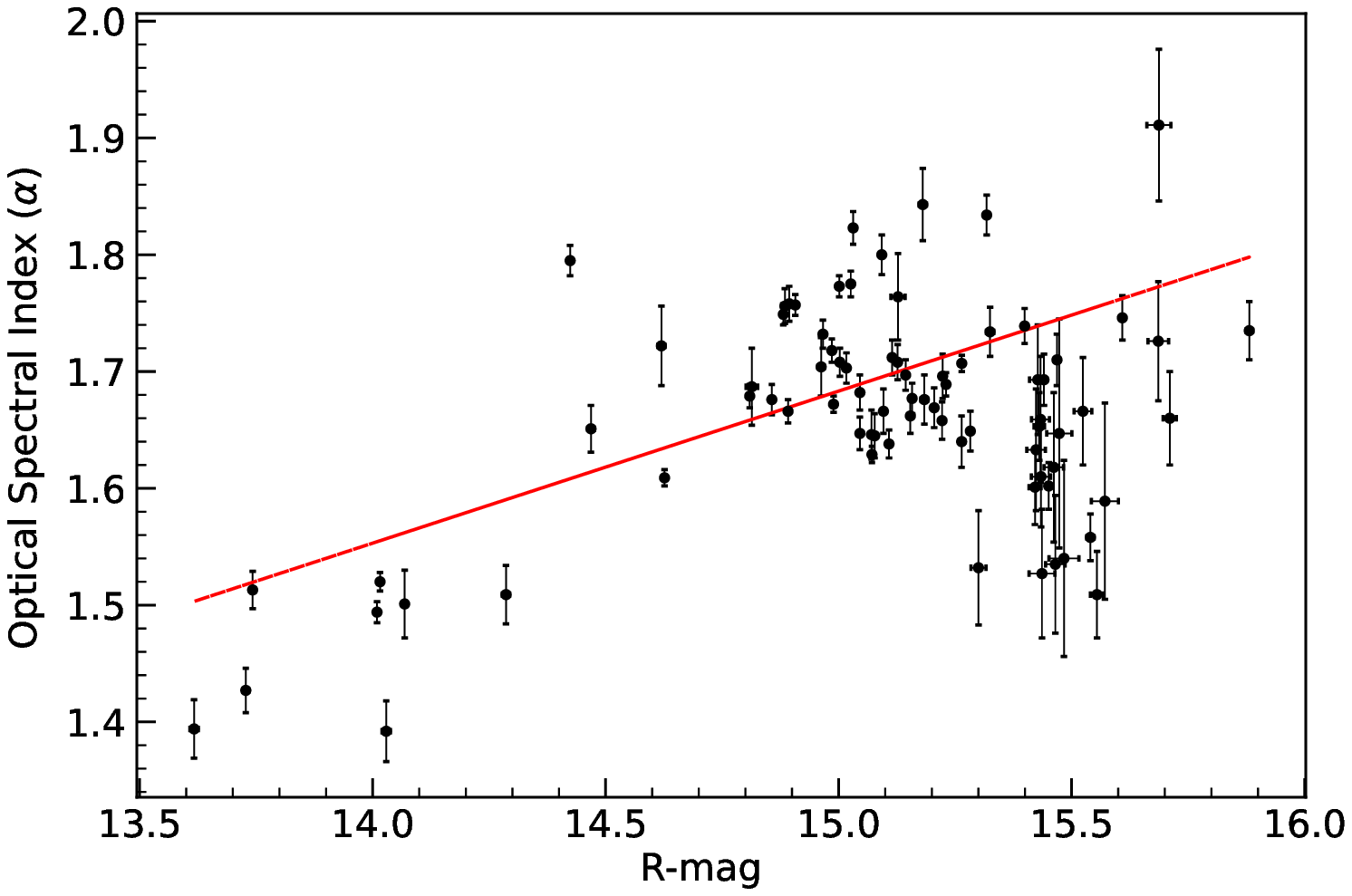}
\caption{Variation of optical spectral index of S5 1803+784 with respect to time (left) and R-magnitude (right) during the entire monitoring period.} 
\label{fig:alpha}
\end{figure}

In the case of blazars, flux changes can be attributed to the color variations, and studying these variations can aid in understanding the underlying emission mechanism.
Two different color trends dominant in blazars are bluer when brighter (BWB) and redder-when-brighter (RWB). BL Lacertae objects generally display a BWB trend while FSRQs show an RWB trend, but in some of the blazars, no clear color behavior has been observed \citep{B_ttcher_2009, 2015MNRAS.451.3882A}.
Moreover, it has also been found that the same source displayed both these behaviors during different flux states \citep{2011MNRAS.418.1640W}. To investigate the color behavior of the blazar S5 1803$+$78, we studied the variation of the spectral index with time and R band magnitude; we plotted the values of the spectral index given in column 2 of Table\,\ref{tab:sed} vs. time (the left one of the Fig.\,\ref{fig:alpha}) and R band magnitudes (the right one of the Fig.\,\ref{fig:alpha}). We fitted straight lines (S =$m$V + $c$) on plots of spectral index, S, against R magnitude and S vs. time.
The values for the slope, $m$, along with the constant, $c$ derived from these fits, are listed in Table\,\ref{tab:alpha_tr}.
A positive slope implies a positive
correlation between the two quantities which signifies that BWB or redder-when-fainter trend is dominant, whereas a negative correlation is observed when we get a negative
slope, which implies an RWB behavior.

\begin{figure}
\centering
\includegraphics[width=8.5cm, height=6cm]{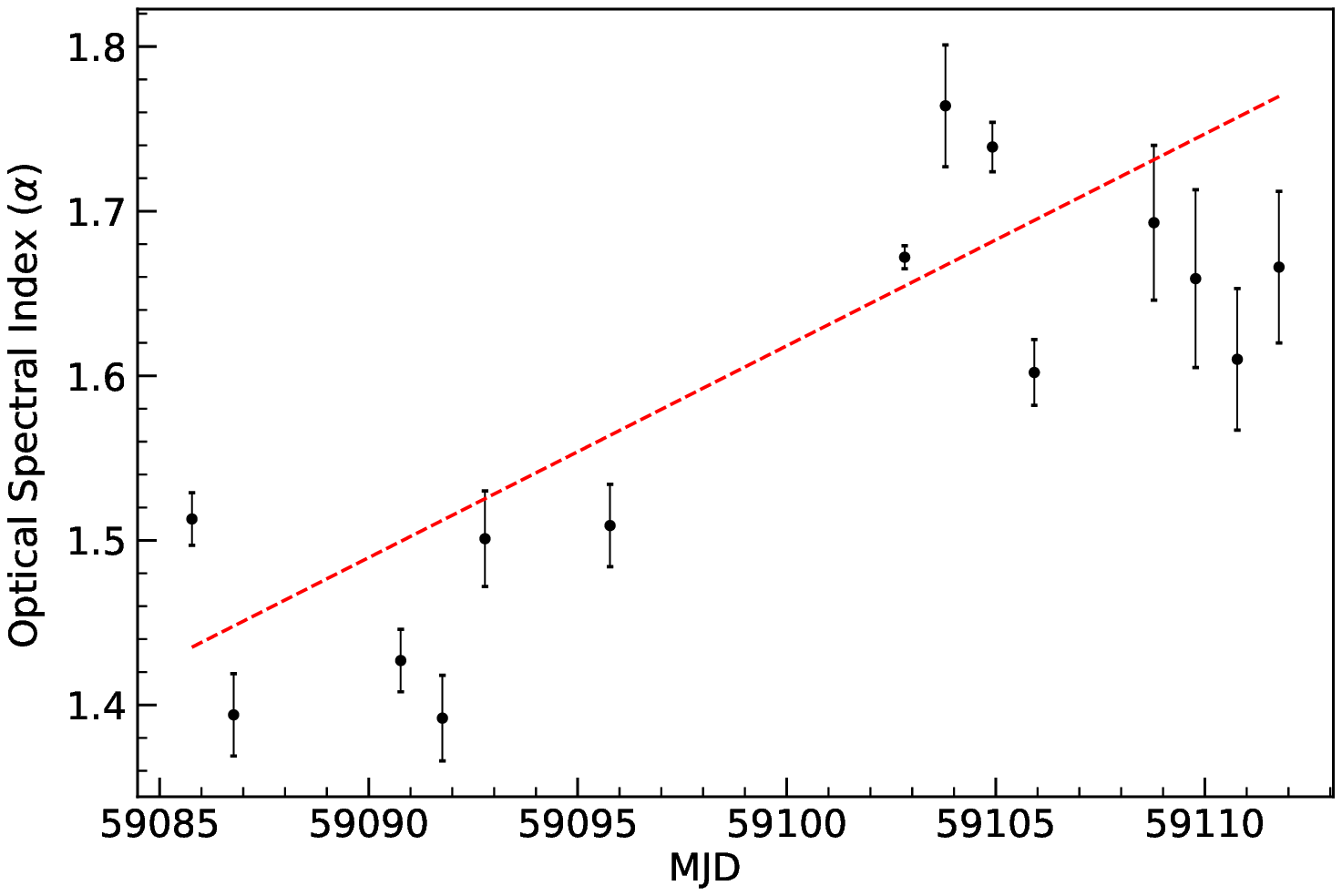}
\includegraphics[width=8.5cm, height=6cm]{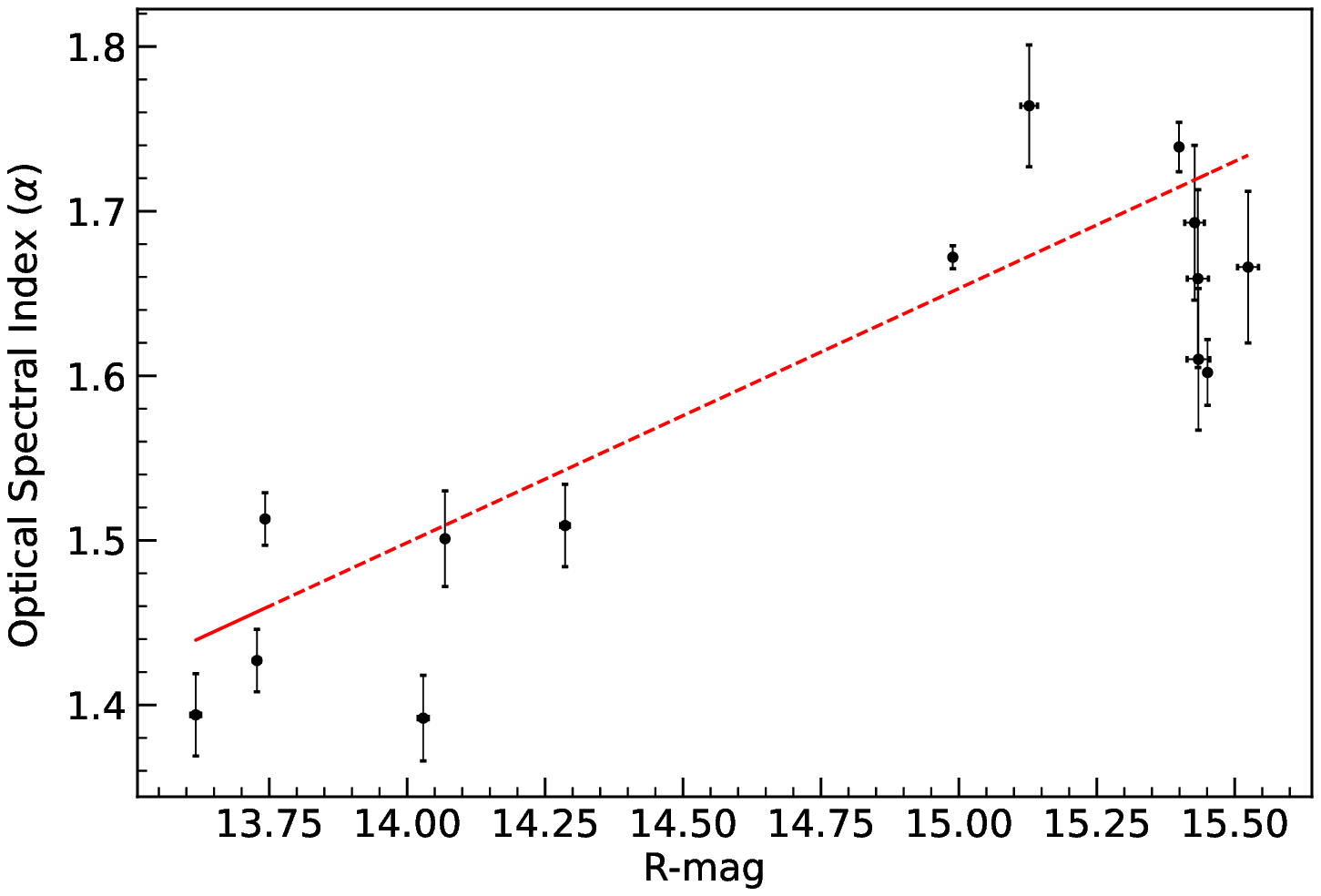}
\caption{Variation of optical spectral index of S5 1803+784 with respect to time (left) and R-magnitude (right) during the flaring period (MJD 59085.77 - MJD 59111.77).} 
\label{fig:alpha_flare}
\end{figure}

A large value of the null hypothesis probability ($p$) implies a higher chance of correlation caused by random noise. In contrast, a smaller p-value suggests a high probability of genuine correlation.
A significant positive correlation (the null
hypothesis probability, $p \le 0.05$) in both the plots indicates that the blazar displays a mild BWB trend on longer timescales; thus, the spectrum of the source hardens as the flux increases. We also examined the variation
of spectral indices with respect to R-band magnitude and time only during the flaring period (MJD 59085.77 - MJD 59111.77)) and found similar trends. The straight-line fits the spectral index vs. R magnitude and time during the flare are displayed in the Fig.\,\ref{fig:alpha_flare} while the results of the straight-line fits are given in Table\,\ref{tab:alpha_tr_flare}. 

\begin{table}
\caption{Variation of spectral index with time and R-mag during the entire monitoring period.} 
\label{tab:alpha_tr} 
\centering 
\resizebox{0.5\textwidth} {!}{
\begin{tabular}{ccccc} 
\hline\hline 
Parameter & $m_2^a$ & $c_2^a$ & $r_2^a$ & $p_2^a$ \\
	& & & & \\ 
\hline 
$\alpha$ vs time  & 3.78e-04$\pm$1.12e-05 & -20.72$\pm$  0.66 &   0.54 & 2.39e-07 \\
$\alpha$ vs R-mag & 1.30e-01$\pm$4.08e-03 &  -0.27$\pm$  0.06 &   0.44 & 4.76e-05 \\
\hline 
\end{tabular}}\\
$^am_2$ = slope and $c_2$ = intercept of $\alpha$ against time or R magnitude; $r_2$ = Correlation coefficient; $p_2$ = null hypothesis probability
\end{table}

\begin{table}
\centering 
\caption{Variation of spectral index with time and R-mag during the flaring period (MJD 59085.77 - MJD 59111.77).} 
\label{tab:alpha_tr_flare} 
\resizebox{0.5\textwidth} {!}{
\begin{tabular}{ccccc} 
\hline\hline 
Parameter & $m_3^a$ & $c_3^a$ & $r_3^a$ & $p_3^a$ \\
	& & & & \\ 
\hline 
$\alpha$ vs time  & 1.29e-02$\pm$6.97e-04 & -759.05$\pm$ 41.21 &   0.81 & 4.33e-04 \\
$\alpha$ vs R-mag & 1.54e-01$\pm$7.94e-03 &  -0.66$\pm$ 0.12   &   0.87 & 5.75e-05 \\
\hline 
\end{tabular}}\\
$^am_3$ = slope and $c_3$ = intercept of $\alpha$ against time or R magnitude; $r_3$ = Correlation coefficient; $p_3$ = null hypothesis probability
\end{table}

\section{Discussion and Conclusion}
\label{sect:disc}
In this paper, we present a set of results from our ongoing campaign to study the optical properties of blazars on diverse timescales. We monitored the blazar S5 1803$+$78 quasi-simultaneously in $BVRI$ on 122 nights to investigate the optical properties of the source.
Our data set includes a total of $\sim$ 2100 $BVRI$ frames collected from May 2020 to July 2021 using three ground-based optical telescopes, i.e., 1.0\,m RC telescope, 60\,cm RC robotic telescope, and 0.5m RC telescope in Turkey.
During this monitoring period, the source brightness in the $R-$band varied from 13.617$\pm$0.009 to 15.888$\pm$0.01. During our observation campaign, we observed the brightest flare for the blazar S5
174 1803+78 till date. The source started flaring around August 2, 2020 and lasted for a duration of 57 days. The blazar was in the brightest ever state on August 25, 2020 with R band magnitude of 13.617.
To investigate source properties in optical regimes during its different brightness states, we performed our analysis on diverse timescales and also during the flare period.
Studying blazars during their outburst state provides an opportunity to understand their variability, spectral behavior, color trends, and dominant emission mechanisms in much greater detail. This unprecedented flare is used to understand various characteristics of the source in detail from minute timescales to yearly timescales. The blazar emission during the outburst state is predominantly understood by relativistic shocks propagating through the jet. In general, during the flaring state non-thermal Doppler-boosted emission from relativistic jet plasma is dominant and is enhanced by the propagation of shocks in the flows \citep{1979ApJ...232...34B}.

Microvariability in blazar sources has gathered special attention, partially because of the requirements of very efficient particle acceleration and very fast energy dissipation mechanisms necessary to produce it.
To study intraday variability during the flaring and non-flaring periods, we observed the source continuously in the $R-$band for $\sim$ 1.5 to 4 hours on a total of 13 nights. To claim the presence of IDV in our source, we used a power-enhanced F-test and the ANOVA test, and only on two instances, the source was found to be variable using both the tests. During the flare period, we studied IDV on three days, i.e., August 30, 31, and September 10, 2020. We did not find any significant variability on either of these nights. The source was variable on minute timescales on only 2 nights with variability amplitude of 7.47 
and 10.10. However, the low duty cycle of the source could be due to the small observation spans. \citet{2005A&A...440..855G} extensively studied microvariability in a sample of AGN classes. One of the major findings is that $\sim$ 60 -- 65\% of blazars are found to be variable on intraday timescales when observed for $<$ 6 hours, whereas this fraction increased to $\sim$ 80 -- 85\% when monitored for more than 8 hours. Therefore, the duty cycle of our source may further increase if we increase the observation duration.
On longer timescales, covering the entire monitoring duration, strong flux variations were seen in all optical bands.

Several detailed works in the literature have been dedicated to studying optical variability in AGNs to understand the location and size of emission region, dominant particle acceleration mechanisms, SMBH mass, and various radiative processes in AGNs. The relativistic jet of blazars is pointed directly towards us, and thus their emission at different frequencies is mainly coming from the extragalactic jet. This makes blazars best sources to study jets, central SMBH, and the accretion disc. Blazar variability is one such tool to understand the blazar structure. Variability could be due to intrinsic or extrinsic mechanisms. According to intrinsic mechanisms, flux variability is widely believed to occur due to particle injection and/or acceleration in a thin section within the jet, such as the jet base or the shock front. Propagation of multiple shocks can cause shock-shock interactions leading to an increase in magnetic field and acceleration of particles. Changes in the magnetic field or electron density can cause flux changes in the source. These weak flux changes are greatly enhanced by the extreme Doppler boosting, thus causing large changes in the observed flux and the timescale of variability. Another possible scenario for the intrinsic mechanism that can cause flux variability to include magnetic reconnection \citep{2019ApJ...887..133B}.
While the extrinsic ones involve geometrical processes (due to the change in the orientation of the emission region with our line of sight), microlensing, and interstellar scintillation \citep{1992A&A...259..109G, 2003ApJ...585..653B, 2017Natur.552..374R}.

Many studies in the past have shown different intraday variability behaviors in various classes of blazars, with BL Lacertae objects being comparatively less variable on the intraday timescales \citep{2011MNRAS.416..101G}.
Fine-scale structures, such as inhomogeneities or bends in the base of jets, can cause intraday variability in the optical light curves of the blazars when interacting with the shocks in jets.
The Kelvin-Helmholtz instabilities responsible for these small scale structures in the relativistic jets of the BL Lacertae objects are prevented if the axial magnetic field exceeds the critical
magnetic field value, which is defined as \citep{1995Ap&SS.234...49R}:
\begin{equation}
B_c = \big[4\pi n_e m_e c^2(\Gamma^2 - 1)\big]^{1/2} \Gamma^{-1},
\end{equation}
where $n_e$ is the density, $m_e$ is the rest mass of the electron, and $\Gamma$ is the jet's bulk Lorentz factor.
The stronger magnetic field of BL Lacertae objects such as our source S5 1803+784 would prevent the occurrence of small-scale instabilities, thus giving less pronounced microvariability.

According to the shock-in-jet scenario, the variability amplitude is larger at higher frequencies.
It has been observed by many authors \citep[e.g.][]{1998MNRAS.299...47M, 2003A&A...397..565P, 2011AJ....141...65D, 2021A&A...645A.137A} and indicates towards the blazar spectra getting steeper with a decrease in the brightness and flatter as brightness increases \citep{1998MNRAS.299...47M, 2003A&A...397..565P}. Whereas, on many instances, the amplitude of variability of blazars at higher wavelengths was either comparable or more than that at lower wavelengths \citep{2000ApJS..127...11G}.
We observed the same trend on longer timescales. The variability amplitude for the blazar S5 1803$+$78 over the entire monitoring period in optical bands B, V, R, I, were found to be 238.481\%, 234.842\%, 227.060\%, and 216.566\%, respectively (Table\,\ref{tab:var_res2}).
Such amplitude trends could be related to the synchrotron mechanisms responsible for optical radiation.
According to energy loss equation (under constant magnetic field) we have $-d\gamma/dt \propto \gamma^2$
where $\gamma$ is the electron Lorentz factor) and $t_{sync} \propto 1/\gamma$, dissipation of a large amount of energy
(by higher energy electrons) as high-frequency photons could happen in shorter timescales.
This, therefore, causes any spectral changes to be observed first in the bluer wavelength followed by redder ones
with a certain delay. Moreover, this amplitude trend can be hint towards the presence of the BWB chromatism in the source on the intraday timescales. 

To further understand the spectral changes in the source, we have generated optical SEDs using the quasi-simultaneous
B, V, R, and I band data points corrected for galactic extinction.
We generated 79 such optical-SEDs and they were found to be well defined by simple power law. We then fitted these SEDs with a straight line to get the spectral indices. Spectral index during our observation period varied from 1.392 to 1.911 while the weighted mean spectral index is 1.673 $\pm$ 0.002. Our results were found to be in good agreement with \citet{1996ApJS..107..541L} and \citet{2021MNRAS.502.6177N}. The steep optical spectra of the source is consistent with synchrotron emission as the dominant process at optical frequencies along with the presence of relatively strong axial magnetic fields of the relativistic jet.

Analysing the variation of spectral indices with time and R band during the whole monitoring duration indicates the presence of a mildly detectable BWB trend ($r\sim0.5$). While during the flare, the source follows a much stronger BWB color behavior ($r\sim0.9$).
Several theories can interpret the mild BWB trend obtained during our long-term observations.
According to \citet{2002MNRAS.329...76H}, the underlying host galaxy effects can cause color changes over time. But with one of their jets pointed towards the observer, the BL Lacertae objects are beamed and intrinsically weak. Therefore, the non-thermal Doppler boosted emission from the relativistic jet powers the observed emission from BL Lacertae objects and swamps the light from the accretion disk, particularly during the active phase. Therefore, the jet-based theoretical models can shed light on the flux and color variability results observed by us. Moreover, as pointed by \citet{2021APh...12902577B}, presence or absence of the BWB trend on diverse timescales can be interpreted by different lifetime of sub-components with various Doppler factors and volumes.

\citet{2014ApJ...792...54S} studied the timescale-dependent color variations. According to their model, the BWB trend is stronger for timescales $<$ 30 days, and it weakens on longer timescales of more than 100 days. Our results are also in agreement with this. As evident from the Tables\,\ref{tab:alpha_tr_flare} and \ref{tab:alpha_tr}, the
correlation coefficient value ($r$) varies from 0.87 (during the flare period of $\sim$ 26 days\footnote{For our color-magnitude analysis during the flare we took a period from MJD 59085.77 to MJD 59111.77 so that we have quasi-simultaneous data points in BVRI passbands.}) to 0.44 (during the entire monitoring period of more than $\sim$ 400 days). 

The correlation between emissions at different frequencies can be used to infer the structure of the blazar, various emission processes at work, and the location of emission regions. It is always difficult to detect time delays in optical wavelengths, due to small wavelength separations among different optical bands. Here, we studied cross-correlation between different optical bands during the brightest flare of the source and also during the entire monitoring period using DCF analysis.
Almost all combinations of optical frequencies in both the cases gave strong correlation with a near-zero time lag which suggest these
emission regions are co-spatial. The flare is expected to start simultaneously in optical frequencies as the optical B, V, R, and I passbands are closely spaced. Therefore, short timescale observations could possibly not detect the time lags among these passbands. We also carried out a periodicity search using four different methods: DACF, SF, WWZ, and LSP. To estimate the significance of the peaks, we followed \citet{1995A&A...300..707T}. No significant periodicity was found on longer timescales spanning a period from May 2020 to July 2021. The unevenly sampled observations make it difficult to search for periodicity in blazars. Any detection of quasi-periodicity can provide important clues on emission mechanisms occurring in the source and also help us in having a better understanding of the various theoretical models for blazars.

This work is part of an ongoing project focusing on understanding the source behavior during the recent unprecedented flare of the source using dense optical observations.
Due to the low cadence of multi-frequency data i.e., X-ray and $\gamma$-ray, understanding source behaviour on minute timescales was not feasible. Therefore, to investigate source behaviour on diverse timescales during different states of the source, we focused on the optical data here.
Near-simultaneous multi-frequency observations play an important role in understanding various problems of blazar physics. In this direction, observations of a large sample of blazars using the 1-2m class telescope facilities can prove to be very useful.
Apart from S5 1803$+$78, we are observing a dozen other $\gamma$-ray blazars using various small aperture telescopes to get important leads in studying the blazar physics and to confirm the findings that already exist.

\section{Acknowledgement}
We thank the anonymous referee for useful comments and
suggestions that helped us in improving our manuscript.
AA and AO were supported by The Scientific and Technological Research Council of Turkey (TUBITAK), through project number 121F427.
We thank TUBITAK National Observatory for partial support in using T60 and T100 telescopes with project numbers 19BT60-1505 and 19AT100-1486, respectively. This research has made use of data obtained using the ATA50 telescope and CCD attached to it, operated by Ataturk University Astrophysics Research and Application Center (ATASAM). Funding for the ATA50 telescope and the attached CCD has been provided by  Scientific Research Projects Coordination Units in Atatürk University (P.No. BAP-2010/40) and Erciyes University (P.No. FBA-11-3283), respectively. AO, EE were supported in part by the Scientific Research Project Coordination Unit of Ataturk University (P. no. FBA-2020-8418). This research has been supported by The Scientific and Technological Research Council of Turkey (TUBITAK) through project number 121F427.

\bibliographystyle{aasjournal}
\bibliography{bibtex}
\end{document}